# An all-optical convolutional neural network for image identification

Wei-Wei Fu[1,2,#], Dong Zhao[1,#], Qing-Hong Rao[1], Heng-Yi Wang[1], Ben-Li Yu[2,3], Zhi-Jia Hu[2,3],

Fang-Wen Sun[1,*], Kun Huang[1,2,*]

[1] Department of Optics and Optical Engineering, University of Science and Technology of China, Hefei, 230026, Anhui, China

[2] State Key Laboratory of Opto-Electronic Information Acquisition and Protection Technology, Anhui University, Hefei, 230601, Anhui, China

[3] School of Optoelectronic Science and Engineering, Anhui University, Hefei, 230601, Anhui, China

[#] *W. F.* and *D. Z.* contributed equally to this work.

[*]Corresponding authors: F. S. (fwsun@ustc.edu.cn), K. H. (huangk17@ustc.edu.cn)

**Abstract**

**In modern artificial intelligence, convolutional neural networks (CNNs) have become a cornerstone for visual and perceptual tasks. However, their implementation on conventional electronic hardware faces fundamental bottlenecks in speed and energy efficiency due to resistive and capacitive losses. Photonic alternatives offer a promising route, yet the difficulty of realizing optical nonlinearities has prevented the realization of all-optical CNNs capable of end-to-end image classification. Here, we demonstrate an all-optical CNN that bypasses the need for explicit optical nonlinear activations. Our architecture comprises a single spatial-differentiation convolutional stage—using 24 directional kernels spanning 360°, along with a mean-filtering kernel—followed by a diffractive fully-connected layer. The directional convolution enhances feature selectivity, suppresses noise and crosstalk, and simplifies the classification task, allowing the weak nonlinearity inherent in optical diffraction to achieve high accuracy. We report experimentally classification accuracies of 86.8% on handwritten digits (MNIST) and 94.8% on a ten-class gesture dataset. The system delivers a computational throughput of $1.13\times10^5$ tera-operations per second (TOPS) and an energy efficiency of $1.51\times10^3$ TOPS/W—the highest reported among CNN hardware—with the potential to improve by a further 5–6 orders of magnitude using nanosecond-scale detectors. This work establishes a scalable pathway toward ultralow-latency, ultralow-energy vision processing for real-time intelligent systems.**

**Introduction**

Convolutional neural networks (CNNs) extract hierarchical features from raw data to reduce parameter complexity and improve prediction accuracy[1,2] in computer vision[3,4], speech recognition[5,6], [6]and medical diagnosis[7,8]. Traditional CNN hardware, such as central processing units [9], graphics processing units [10,11], and application-specific integrated circuits [12-14], operate on the von Neumann architecture with synchronous digital clocking. Accordingly, their computing speed is primarily limited by clock frequency and memory access time[15], because electronic components face fundamental constraints from Joule heating, electromagnetic crosstalk, and slow capacitor charging/discharging in semiconductor materials[16,17]. These inherent limitations restrict further efficiency gains, driving the need for alternative computing paradigms.

Photonic computing is a potential substitute for electron-based computing, data transfer and storage due to its light-speed parallel operation with high energy efficiency and large throughput[18-27]. By implementing time-cost convolution or multiplication operations via optical approaches, such as diffraction[28-34], interference[35-37], wavelength-division multiplexing[38-42] and tunable optical attenuation[43,44], optoelectronic CNNs accomplish ultimate identification tasks with digital neural networks, demonstrating performance comparable to the state-of-the-art digital processors. Nevertheless, these optoelectronic CNNs require extensive analog-to-digital converters for digitization, where high-throughput and high-precision data transmission introduces significant energy consumption and latency. Additionally, their on-chip implementations[34-36,38-40] often reduce two-dimensional images to one-dimensional input vectors, thus limiting input capacity due to finite photonic integration footprint, evitable waveguide crossings and thermal crosstalk.

In principle, all-optical CNNs (AOCNNs) can overcome these aforementioned limitations due to superior processing speed and energy efficiency for all-optical signals in physical neural networks[45-47]. However, AOCNNs for image identification remain unrealized due to difficulties in implementing effective optical nonlinear activation. Linear optical models exhibit limited recognition performance on complex datasets and multi-class tasks[48,49]. Although approaches such as carrier effects micro-ring resonators[50,51], phase transition of material[52,53], and multiple scattering of light[54,55] have been explored, achieved nonlinearity remains insufficient for full operation of AOCNNs, thus necessitating additional digital components to accomplish the ultimate tasks.

Here, we present a simplified AOCNN architecture comprising only a convolutional layer and a fully-connected layer (FCL), which performs high-accuracy image identification without additional nonlinear activation components. By enhancing inter-class orthogonality and suppressing signal crosstalk through directional differentiation in the convolutional stage, the system effectively leverages weak optical nonlinearity introduced via a tailored random diffractive phase in the FCL to boost recognition performance. Without changing the convolutional unit, the diffractive phase of the FCL can be dynamically reconfigured using a spatial light modulator (SLM) to adapt to different classification tasks. Experimentally, the system achieves accuracies of 86.8% on MNIST and 94.8% on a ten-class gesture recognition dataset, while delivering a computational speed of $1.134 \times 10^5$ TOPS and an energy efficiency of $1.508 \times 10^3$ TOPS/W—surpassing all previously reported CNN implementations. As computation occurs instantaneously during light propagation, these results are achieved without any electro-optic conversion bottlenecks and can be further improved by 5–6 orders of magnitude using nanosecond-scale photodetection.

**Architecture of AOCNN**

Digital CNNs typically comprise multiple fundamental components—such as convolutional layers, nonlinear activation functions, pooling layers, and FCLs. In AOCNNs, however, pooling layers—commonly employed in digital systems to reduce spatial dimensions and compress features—become redundant. This simplification stems from the inherent parallelism of optical computing: photons perform operations simultaneously at the speed of light within the network, rendering computational complexity relevant only to system optimization rather than actual inference latency. Therefore, we introduce a highly streamlined AOCNN architecture that retains only the essential convolutional and fully-connected layers, enabling high-speed and energy-efficient image identification (Fig. 1a). In this framework, nonlinear operations are intrinsically achieved via optical scattering derived from the diffractive phase profiles in the FCL.

The optical convolutional layer is implemented within a thin-lens imaging system configured at unity magnification, wherein both the input and output planes are positioned at a distance of $2f$ from the lens. A multi-kernel meta-modulator is situated at the back-focal plane (Fig. 1a), enabling arbitrary customization of the system's point-spread function (PSF). Inspired by the edge-primacy mechanisms observed in biological[56] and computer visions[57], the meta-modulator employs complex-amplitude modulation, denoted as $h_{\text{meta}}$, to generate a predefined 5×5-array PSF (see distribution in Fig. 1b and detailed parameters in Methods). This structured PSF comprises 24 rotated Sobel operators and one mean-filtering operator, collectively facilitating the extraction of directional edge features from an input image through selected differentiations across 360° (Fig. 1b). According to wave optics[58], the complex amplitude profile of the meta-modulator is derived from the inverse Fourier transform of the target PSF (see Supplementary Section 1).

The arrayed configuration of Sobel operators enables parallel convolutional processing, thereby supporting high-speed optical computation.

The directional differentiation implemented in the convolutional layer serves to amplify distinctions between classified images. For example, digits '6' and '8' differentiated along $\theta = 0°$ (Fig. 1c) exhibit substantially reduced spatial overlap compared to their original forms. To quantitatively assess this effect, we compute the structural similarity index (SSIM; defined in Supplementary Section 2) between the two digits using 800 randomly selected image pairs from the MNIST dataset. The resulting SSIM values follow a Gaussian distribution (Fig. 1d), with the peak probability after differentiation shifting toward significantly lower values relative to the originals—indicating enhanced orthogonality and improved discriminability between the processed digits. This improvement is consistently observed across all digit pairs, as evidenced by the reduced peak SSIM values for each combination (Fig. 1e). Moreover, across differentiation angles from $\theta = 0°$ to $180°$, the average SSIM for all digit pairs remains consistently around one-seventh of that of the undifferentiated images, demonstrating robust performance independent of differentiation direction. Owing to these advantages, the convolutional layer is kept fixed throughout this study, enabling adaptation to various image identification tasks by modifying only the optical fully-connected layer.

The optical FCL is implemented using a single engineered diffractive element—for instance, a reconfigurable spatial light modulator (SLM) as depicted in Fig. 1a. This component enables dynamic adjustment of neural network biases and weights via an encoded quasi-random phase profile. Within this FCL, light conveying the convolved image array undergoes optical scattering—a recently demonstrated mechanism for introducing weak optical nonlinearity through diffractive phase modulation[54,55]. Upon transmission through the FCL, image identification is completed via square-law detection of the output intensity across each detector region (Fig. 1a), effectively implementing second-order nonlinear processing. Although both nonlinearities induced by scattering and square power in passive manners are weaker than numerical nonlinearity, our specially designed convolutional layer for extracting discriminative directional features of an image can compensate for this nonlinearity limitation.

**Performances of AOCNN**

To showcase the capabilities of our approach in background noise suppression, signal crosstalk reduction, accuracy improvement, and scalability, we numerically simulate an AOCNN for handwritten digit recognition using the MNIST dataset, see its network architecture parameters in Supplementary Fig. 1 and Methods. Since the convolutional layer remains fixed, training is performed by optimizing only the phase profile of the FCL via an Adam-gradient descent algorithm within a TensorFlow framework[59] (see Methods for optimization specifics). For comparison, we also construct a control architecture by removing the convolutional layer from the AOCNN, resulting in a single-layer diffractive neural network (DNN; see Supplementary Figs. 2a and 2b and Supplementary Section 3)[46]. This baseline model serves to evaluate the relative performance of the full AOCNN system.

**Signal-to-noise ratio (SNR) enhancement**. After training for $10^4$ epochs on the MNIST training dataset using 6,000 randomly sampled digits spanning all ten classes, we evaluate the performance of both the AOCNN and the single-layer DNN by identifying a sample digit '6' from the test set and comparing their signal-to-noise ratios (SNRs). The AOCNN exhibits markedly lower background intensity at the detector plane (Fig. 2a) relative to the DNN (Fig. 2b), as reflected in the output intensity profiles across all ten digit classes. To quantitatively assess background suppression, we analyze 800 randomly selected handwritten '6' samples from the MNIST test set, computing the SNR for each—defined as the ratio of optical power inside the designated detector region for digit 6 to that outside it. Statistical analysis of the resulting 800 SNRs (Fig. 2c) confirms that the AOCNN achieves significantly higher SNR than the DNN, underscoring its superior noise suppression and signal clarity. Furthermore, comparative evaluation across all digit categories (Fig. 2d) shows that the AOCNN attains an average SNR improvement of more than 5 dB over the DNN.

**Crosstalk reduction.** To quantify inter-channel crosstalk, we evaluate the orthogonality among the ten identification channels by analyzing the normalized optical power distribution across the ten detector regions for each input digit. Crosstalk (denoted as $C$) is defined as the ratio of the maximum energy detected in any non-target channel (i.e., off-diagonal elements in Fig. 2e) to the energy in its corresponding correct target channel (Fig. 2e). According to this metric, correct recognition occurs when $C < 1$, whereas $C \geq 1$ indicates a misclassification. When tested with 800 samples of digit '6' from the MNIST test set, the AOCNN achieves a higher correct prediction rate than the DNN (Fig. 2f). This advantage is consistent across all ten digit categories, with the AOCNN showing a greater average accuracy in statistical comparisons (Fig. 2g). These results align well with the theoretically predicted accuracy for a single FCL, as shown in Fig. 2h.

**Accuracy improvement.** When utilizing a single FCL configuration, the AOCNN demonstrates significantly superior recognition performance compared to conventional DNN, owing to its specially designed convolutional layer that minimizes inter-class crosstalk among output detector areas. For FCL numbers greater than 1, both accuracies for AOCNN and DNN are not improved significantly, compared with that of our single-FCL AOCNN (Fig. 2h). This result implies that the single-FCL AOCNN offers sufficient capability for image identification, thereby avoiding alignment errors in multi-layer DNNs[46]. Compared with single-layer or multi-layer DNNs, our AOCNN exhibits a higher SNR (see Supplementary Fig. 3 and Supplementary Section 4) to maintain robust identification capabilities.

**Scalability.** To further improve the accuracy of the AOCNN, it is essential to minimize the optimized loss function by increasing either convolutional kernels or neurons in the optical fully-connected layer (FCL). Under a fixed dataset size, the number of model parameters substantially influences performance[60]. Although expanding the number of convolutional kernels can improve accuracy, such improvements saturate and yield diminishing returns when the number of kernels exceeds 25, as illustrated in Supplementary Fig. 4a and Supplementary Section 4. Notably, the AOCNN demonstrates good scalability, a trait also observed in conventional digital CNNs. As the input dimension $N$ increases and the pixel size (PS) decreases proportionally, the nonlinear effects induced by optical scattering become more pronounced. Consequently, the optimized loss function exhibits a power-law dependence on $N^2$ (Fig. 2i), described by the relation: $Loss=[N^2/(3.58\times10^6)]^{-0.83}$. This scaling facilitates a substantial enhancement in classification accuracy, which exceeds 95% for a pixel size of 0.93 μm (Fig. 2j), suggesting that the AOCNN's accuracy can closely match that reported for single-layer digital CNNs (98.96% in Ref. [61]).

**Experimental verification of AOCNN on MNIST dataset**

To experimentally validate the multi-kernel meta-modulator—whose complex amplitude profile is shown in Fig. 3a—we employ transmissive dielectric geometric metasurfaces composed of size-and rotation-tunable nanobricks (Fig. 3b) [62,63]. Within this design, the in-plane dimensions and rotation angles of the nanobricks independently govern amplitude and phase modulation[58], respectively. Detailed design and characterization procedures are provided in Methods. Under illumination at $\lambda$ = 635 nm, we detected the edge of an amplitude object, illustrated by the digit '3' in Fig. 3c, using a self-made optical setup (see Supplementary Fig. 6a). The meta-modulator selectively enhances edges along the intended direction (Fig. 3d), with close agreement between simulation and experiment (see Supplementary Figs. 6b and 6c) confirming the validity of the edge detection.

After the input images undergo processing through the fixed convolutional layer, the phase profile of the FCL is optimized via a forward propagation and error backpropagation algorithm (see Supplementary Fig. 1b), where optical diffraction is numerically simulated using angular spectrum method[64]. When trained for $10^4$ epochs on preprocessed MNIST dataset (see Methods), the ACONN architecture achieves a recognition accuracy of approximately 90% (Fig. 3e) with only a single optical FCL, demonstrating its high-precision classification capability. The optimized phase (Fig. 3f), along with an incorporated lens phase (Fig. 3g) that enhances energy convergence toward the detector area, is subsequently mapped onto a two-dimensional SLM (Fig. 3h).

To experimentally characterize the performance of the AOCNN, we input individual test digits from '0' to '9' from the MNIST dataset into a custom AOCNN setup (see Supplementary Fig. 7a). Input light representing each digit is directed toward ten predefined regions on the output plane, with the detector region receiving the highest optical energy interpreted as the classification result. For example, for the handwritten digit '0' (Fig. 3i), the AOCNN successfully concentrated the optical energy from the input source into the detector region corresponding to digit '0', as shown in both simulation (Fig. 3j) and experimental (Fig. 3k) intensity profiles. Quantitatively, the measured optical power is the highest at detector '0' (Fig. 3l). Additional recognition examples are provided in Supplementary Fig. 7b.

When evaluating 1,000 images (100 per digit) randomly selected from the MNIST test dataset, the AOCNN achieved a classification accuracy of 89.4% in simulation (Fig. 3m) and 86.6% experimentally (Fig. 3n), with stable performance maintained over 600 minutes (Fig. 3o). The slight discrepancy between simulation and experiment is attributed to fabrication imperfections and misalignments between the convolutional layer and the optical fully connected layer, which could be mitigated through adaptive training[24]. Furthermore, we observe that the classification performance of the AOCNN is inversely correlated with the complexity of the input categories, reaching perfect classification accuracy (100%) in a two-class scenario (Fig. 3p).

**Reconfigurable AOCNN for gesture recognition**

Having established the efficacy of our AOCNN on handwritten digit recognition, we next explore its reconfigurability by adapting the system to classify ten hand gestures (Fig. 4a; see Methods for details) under more practical conditions. This reconfiguration requires only the optimization of the optical FCL (see Supplementary Figs. 8a and 8b for the corresponding loss function, accuracy and phase map), followed by an update to the phase pattern displayed on SLM. Gesture images are encoded as amplitude inputs to the AOCNN for network training. The output plane is divided into ten detector regions—identical in layout to those used in digit recognition—each assigned to a specific gesture category. As an example, for an input representing gesture '6' (Fig. 4b), both simulated (Fig. 4c) and experimental (Fig. 4d) output patterns show the highest optical intensity (Fig. 4e) at the detector corresponding to gesture '6', thus confirming successful classification. Comparable performance is also observed across other gesture categories (see Supplementary

Fig. 9).

On a test set of 1,000 gesture samples collected from four volunteers, the AOCNN achieves a classification accuracy of 94.8% (Fig. 4f), slightly below the simulated value of 98.8% (see Supplementary Fig. 8c). To evaluate practical applicability, we examine the system's robustness to gesture rotation by feeding rotated inputs into the network. Experimental measurements show that crosstalk remains below the correct-recognition threshold of 1 across a broad range of rotation angles (Fig. 4g), indicating strong rotational robustness. Across all gesture categories, the average rotation tolerance is approximately 42° (Fig. 4h), making the AOCNN suitable for real-time operation in dynamically changing environments. To validate it, an experimental demonstration of real-time recognition of the ten gestures under laboratory conditions is provided in Supplementary Video 1, with full details described in Supplementary Fig. 10 and Supplementary Section 6.

**Discussions**

The PSF array employed here is designed through a physics-guided model to enhance image detail contrast, and remains fixed across diverse identification tasks. For specialized applications beyond image recognition, both the PSF array and the FCL can be co-optimized during training, resulting in improved performance such as higher classification accuracy and larger SNR. Beyond factors including the number of convolution kernels and class categories (see Supplementary Fig. 4a and Supplementary Section 4), recognition accuracy is also affected by the spatial dimensions of individual detector regions. Simulations (see Supplementary Fig. 4b and Supplementary Section 4) indicate that reducing detector area significantly improves recognition precision, leading to the selection of an optimal detector size of 18.7 μm × 18.7 μm in our system.

By performing all computations optically, our AOCNN achieves a computational throughput of $1.134 \times 10^5$ TOPS and an energy efficiency of $1.508 \times 10^3$ TOPS/W (see Supplementary Section 7)—values limited only by the frame rate of the recording CMOS camera. Nevertheless, our system outperforms all previously reported CNNs (including those based on all-electronic chips[65]) in both computational speed and energy efficiency (Fig. 4i), while maintaining competitive accuracy (see Table S1). Replacing the current detector with one featuring nanosecond-scale response time[31] could further improve computational speed and energy efficiency by five to six orders of magnitude.

The AOCNN exhibits distinct advantages over other all-optical neural networks, including high stability,

reconfigurability, and low crosstalk. Its single-diffractive-layer architecture eliminates the need for precise alignment between multiple layers and mitigates mechanical instability issues inherent in multi-layer diffractive networks[46]. Furthermore, compared to on-chip all-optical neural networks based on Mach–Zehnder interferometer arrays[45], the AOCNN enables more efficient two-dimensional image input—free from input capacity constraints—while offering reduced crosstalk and enhanced compatibility with conventional digital CNNs.

In summary, we have experimentally demonstrated a reconfigurable AOCNN that utilizes directional differentiation and weak diffraction-induced nonlinearity for image identification under low-power laser illumination. By replacing the current detector with high-speed photodetectors, the AOCNN is expected to achieve computational speed and energy efficiency improvements of several orders of magnitude, far surpassing all existing CNN processors. This architecture opens a pathway to ultra-high-throughput, low-power processing of large-scale image data, supporting real-time machine vision systems and enabling next-generation artificial intelligence hardware.

**Methods**

**Numerical simulation of the AOCNN**

The AOCNN is implemented within a TensorFlow framework. Optical diffraction from the initial input image to the target detector plane is numerically simulated using an angular spectrum method, with a fixed sampling array of 1240×1240 pixels and a pixel pitch of 3.74 μm × 3.74 μm, matching the spatial resolution of our experimental SLM (Holoeye GAEA-2.1). In our simulations, each image to be identified is resized to 228 × 228 pixels and then expanded into a 1240 × 1240 matrix via uniform zero-padding around its periphery, thereby forming the input image for the AOCNN.

As illustrated in Fig. 1a, the input image is positioned at a distance of $2f$ from a thin spherical lens approximated by a phase profile of $\exp(-0.5i \cdot kr^2/f)$, where $k=2\pi/\lambda$, the operating wavelength $\lambda=635$ nm is used here, $r$ denotes the radial position of the lens, and the focal length is $f=50$ mm in this work. Under plane-wave illumination, the diffracted light is collected and projected by the lens onto an imaging plane—also located $2f$ from the lens—after undergoing complex-amplitude modulation via a metasurface-based 25-kernel operator (detailed design provided later in Methods). At this imaging plane, a processed image array of size 1240 × 1240 is generated as the output of the convolutional layer. Due to its co-location with the FCL (i.e., $d=0$ in Fig. 1a), the processed image is immediately diffracted by the FCL phase profile for classification. For each image, the FCL phase is optimized (as described subsequently) to concentrate optical power within a predefined region at the target plane, located $L = 85$ mm from the FCL. A complete mathematical description of the AOCNN is provided in Supplementary Fig. 1 and Supplementary Section 1.

**Design and fabrication of multi-kernel meta-modulator**

To implement 5×5-kernel convolution in the coordinate domain, we employ a metasurface-based complex modulator positioned at the rear focal plane (i.e., the spatial-frequency domain) of the lens in Fig. 1a. This configuration enables the acquisition of a multi-functional PSF, as illustrated in Fig. 1b, for parallel image processing. The designed PSF incorporates 24 directional differentiation sub-operators and one mean-filtering sub-operator, leveraging the superior discrimination capability of differentiation operations in image analysis. Considering the diffraction limit of our imaging system, the spatial resolution for each sub-operator is set to $w_0 = 4p_{SLM}$, where $p_{SLM} = 3.74$ μm corresponds to the pixel size of the SLM. The differentiation sub-operators are generated by rotating a 3×3 Sobel operator (inset, Fig. 1b) from 0° to 180° with an angle interval of 7.5°. Adjacent sub-operators are separated by $228p_{SLM}$ along the $x$- or $y$-direction to align with the input image dimensions. Regions outside these sub-operators are set to zero, resulting in a PSF matrix of size 1140×1140. To satisfy numerical sampling requirements, the PSF matrix is expanded to 1240×1240 by symmetric zero-padding along its periphery.

The complex amplitude of the meta-modulator is obtained via optical Fourier transform (FT) of the ideal PSF in Fig. 1b, following a mathematical framework established in our previous work[58]. In this process, the ideal PSF is placed at the front focal plane of the lens and treated as the source field. Propagation through the lens generates the spatial-frequency spectrum at the rear focal plane, where entire diffractive process is simulated using the angular spectrum method. The resulting field, normalized to its maximum amplitude, defines the complex amplitude (see Fig. 3a) of the multi-kernel meta-modulator. By accounting for the spatial resolution constraints of the optical system, this approach inherently excludes spatial frequencies beyond the lens processing capability.

We implement the modulator using geometric dielectric metasurfaces, which provide independent control over amplitude and phase[58]. These metasurfaces convert circularly polarized incident light into its cross-polarized state, imparting a phase shift equal to twice the rotation angle $\theta$ of the nanobrick, while the conversion efficiency (related with the amplitude modulation) is governed by the nanobrick dimensions (see Supplementary Section 5). Since the rotation and size of each nanobrick can be adjusted independently, phase and amplitude of the cross-polarized transmission are decoupled. This allows mapping the desired complex amplitude of the meta-modulator to specific nanobrick rotation angles and sizes via a pre-calculated lookup table.

Experimentally, the geometric dielectric metasurfaces were fabricated on a 300 nm-thick crystalline silicon (c-Si) film on a sapphire substrate. A positive electron-beam resist (AR-P 6200) was spin-coated and baked onto the silicon layer, followed by patterning via electron-beam lithography (JEOL JBX 6300FS, 100 kV). After development, the sample was rebaked to remove residual moisture. A 10 nm-thick chromium hard mask was deposited via electron-beam evaporation (Kurt J. Lesker PVD75 Proline), and lift-off was performed. Exposed silicon regions were etched using an inductively coupled plasma etcher (Oxford Plasma Pro System100 ICP380), and the remaining chromium was removed with a chromium etchant. Scanning electron microscopy confirmed the high quality of the fabricated meta-modulator (Fig. 3b).

As shown in Supplementary Fig. 5a, the simulated and measured amplitude profiles exhibit a peak near $L = 160$ nm, indicating that the rising edge of this resonance can be utilized for amplitude modulation. The meta-modulator in this work employs seven discrete amplitude levels: 0, 0.05, 0.1, 0.3, 0.6, 0.85, and 1, achieved via nanobricks of dimensions ($W = 90$ nm, $L = 90$ nm), (90 nm, 100 nm), (90 nm, 110 nm), (90 nm, 120 nm), (90 nm, 130 nm), (90 nm, 140 nm), and (90 nm, 160 nm), respectively. Although this discretization may introduce errors, high-fidelity 256-level phase modulation (see Supplementary Fig. 5b) is employed to mitigate inaccuracies and preserve overall performance.

**Experimental performance of optical convolutional layer**

To evaluate the optical performance of the proposed multi-kernel meta-modulator in performing convolutional operations, we construct an experimental setup as illustrated in Supplementary Fig. 6a, including light-source unit, image input unit, optical convolutional layer and detection system. A 635 nm laser beam is first expanded and collimated using a pair of lenses ($L_1$ and $L_2$), followed by linear polarization via a polarizer ($P_1$) to match the input requirements of an amplitude-type spatial light modulator ($SLM_1$, HES 6001, HOLOEYE). The amplitude-modulated beam reflected from SLM1 is then directed by a beam splitter ($BS_1$) and converted into circularly polarized light using a combination of a linear polarizer ($P_2$) and a quarter-wave plate ($QWP_1$). This beam is subsequently focused by a lens ($L_3$) onto our meta-modulator, which generates the convolutional pattern at the image plane. The resulting intensity distribution of the convolved image is captured by a camera. To suppress background noise with the same handedness as the input beam, a crossed circular polarizer assembly—comprising a quarter-wave plate ($QWP_2$) and a linear polarizer ($P_3$)—is placed in front of the camera.

Figure 3d displays the camera-captured image corresponding to an exemplified input digit "3" (Fig. 3c), revealing the anticipated directional operations along with a reconstructed pattern at the center. To quantitatively assess the performance, we compare the simulated and experimentally measured image arrays in Supplementary Figs. 6b and 6c. The line-scanning intensity profiles derived from both datasets show strong agreement, thereby validating the effectiveness of the multi-kernel meta-modulator in executing array convolution.

**Training FCL in AOCNN**

In the proposed AOCNN, the convolutional layer is fixed; only the phase profile of the FCL is optimized using the Adam gradient descent algorithm within the Python version 3.9.13. and TensorFlow framework version 2.5.0 (Google Inc.). The mathematical foundation for training the FCL phase is detailed in Supplementary Section 1. A flowchart illustrating the phase optimization process is provided in Supplementary Fig. 1b for clarity. All training is performed on a workstation equipped with an Intel Core i7-13700KF CPU (@3.40 GHz), 192 GB of RAM, and a GeForce GTX 4090 GPU for accelerated computation.

For the training about digit recognition, we use the MNIST dataset[66], which consists of 60,000 training and 10,000 test samples of handwritten digits across 10 classes. Each sample is binarized, resized to 228×228 pixels, and zero-padded to match the input dimensions of the optical system.

To evaluate performance on gesture recognition, we construct a custom dataset based on the EgoGesture dataset[67]. It comprises 10 gesture classes, corresponding to digits '0' through '9' (Fig. 4a), collected from eight volunteers. Data from four volunteers are assigned to the training set (12,000 images in total, with 300 examples per gesture per individual), and the remaining four participants formed the test set (1,000 images, with 25 samples per gesture per individual). All gesture images are similarly resized to 228×228 pixels and binarized before being used as amplitude inputs.

**Experimental setup of AOCNN**

To experimentally validate the AOCNN, we construct an optical setup to evaluate its identification accuracy. As illustrated in Supplementary Fig. 7a, the setup consists of five main components: a light-source unit, an image input unit, an optical convolutional layer, a FCL, and a detection unit. The first three units—light source, image input, and optical convolution—are identical to those presented in Supplementary Fig. 6a. The FCL incorporates a phase-type $SLM_2$, which encodes a trained phase profile along with an additional lens phase, enabling optical power to be directed to the corresponding detector region for each identified image. An aperture ($AP_2$) is employed to selectively transmit the desired optical signal by filtering out unwanted

zero- and higher-order diffraction generated by the $SLM_2$. The intensity distribution at the detector plane is collected using a lens ($L_4$), which projects the optical pattern directly onto a camera for recognition. Supplementary Fig. 7b compares the simulated and experimentally measured intensity profiles for digits '1' to '9', demonstrating close agreement and confirming accurate digit identification.

Owing to the high reconfigurability of the SLM, the same optical setup can be adapted to perform other recognition tasks—such as the gesture identification presented in Fig. 4—by simply updating the trained phase profile. Supplementary Fig. 9 displays the experimentally obtained intensity distributions for each gesture, showing strong consistency with the corresponding simulations.

**Author Contributions**

K. H. conceived the idea. W. F. and Q. R. developed the theory and carried out the simulations. W. F., B. Y. and Z. H. designed the samples. D. Z. and H. W. prepared and fabricated samples. W. F., K. H. and F. S. implemented the experimental measurement. W. F., K. H. and F. S. wrote the manuscript with inputs from other authors. K. H. and F. S. supervised the overall work. All the authors discussed the results and commented on the manuscript.

**Acknowledgements**

The authors thank the National Natural Science Foundation of China (Grant Nos. 62322512, 62225506, 62505308 and 12134013), the Fundamental Research Funds for the Central Universities (WK2030000108), CAS Project for Young Scientists in Basic Research (Grant No.YSBR-049) and the support from the University of Science and Technology of China's Center for Micro and Nanoscale Research and Fabrication. D.Z. thanks the China Postdoctoral Science Foundation under Grant Number 2023M743364 and "the Fundamental Research Funds for the Central Universities" under Grant Number WK2030000090. W. F. thanks the Anhui Science and Technology Department (Grant No. 2508085QF246). The numerical calculations were partially performed on the supercomputing system at Hefei Advanced Computing Center and the Supercomputing Center of the University of Science and Technology of China.

**Conflict of interest**

The authors Q. R., H. W., B. Y., Z. H and F.S. declare no competing financial interests. The authors K. H., W.F., and D.Z. declare the following competing interests. K. H., W. F., and D. Z. have filed one patent application related to this work through the University of Science and Technology of China. This patent (K. H., W. F., and D. Z., "A design method of all-optical convolutional neural network for image identification", patent No. ZL202510057557.7(2025)) has been granted. This patent applied by University of Science and

Technology of China refers to the design method and physical architecture of all-optical convolutional neural network and its corresponding demonstration for image identification.

**Figures and Captions**

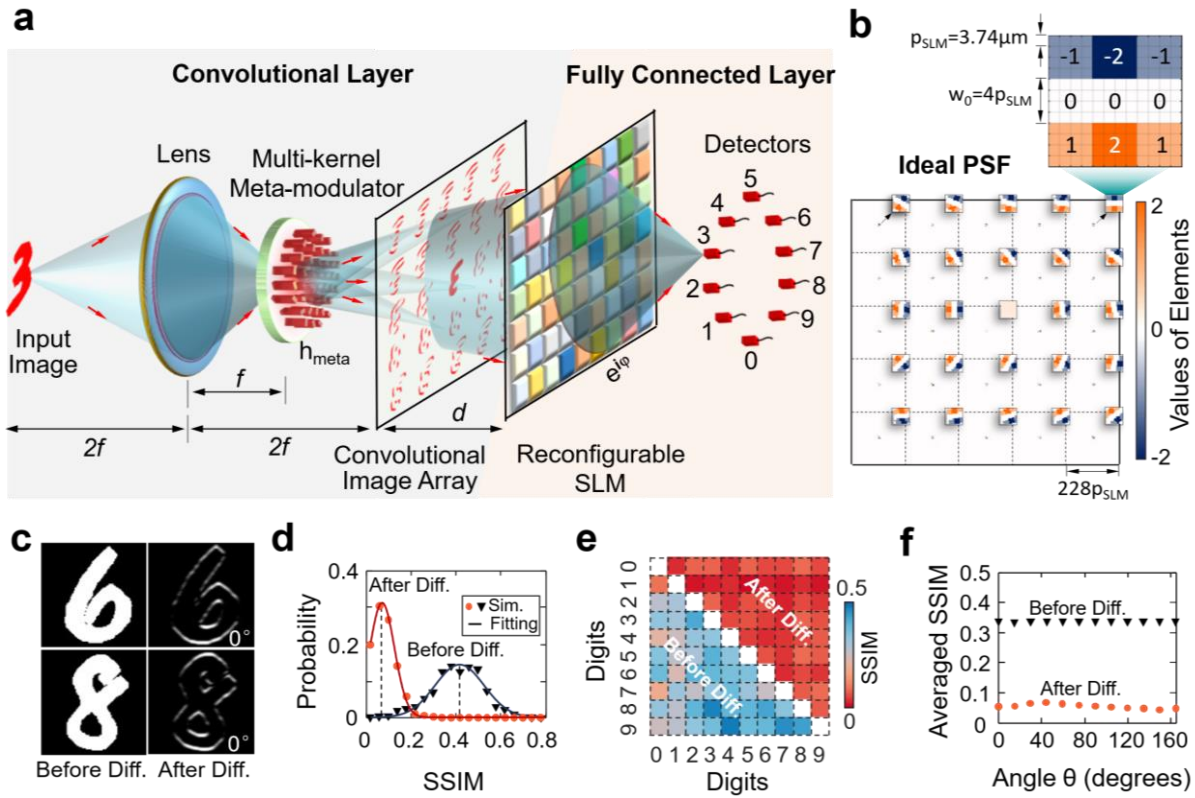

**Figure 1 | Working principle of AOCNN. a,** Sketch for the AOCNN integrating a convolutional layer and a FCL. The convolutional layer consists of a lens with a focal length of 50 mm and a meta-modulator designed to perform multi-kernel convolution operations. The digital image '3' is exemplified here to show its recognition by projecting the maximum energy on its predefined detector. **b,** Ideal PSF designed with 24 directional Sobel operators and a mean-filter operator. Each convolution kernel comprises matrix elements with a spatial resolution of $w_0 \times w_0$, where $w_0$ is set to $4p_{SLM}$ and $p_{SLM} = 3.74$ μm denotes the pixel size of the SLM. A magnified view of the Sobel operator rotated by 0° is provided in the inset, illustrating the dimensional relationship between $w_0$ and $p_{SLM}$. Each kernel is displayed in a zoomed-in format at the upper-right corner of its corresponding operator, thus facilitating observation of each rotated Sobel operator. **c,** Images of digits '6' and '8' before (left) and after (right) convolution with the Sobel operator $G_y$ that can extract horizontal edges. **d,** Simulated and fitted probability density function before and after convolution with the operator $G_y$. 800 image pairs of digits '6' and '8' are randomly selected from the MNIST dataset. **e,** SSIMs at peak probability (see dashed lines in **(d)**) between different digit pairs before (left-bottom) and after (right-top) convolution for horizontal differentiation. **f,** Averaged SSIMs of all digit pairs for each differentiation (solid dots) with the angle ranging from 0° to 180°. The control case (triangles) without differentiation operation is also shown for comparison.

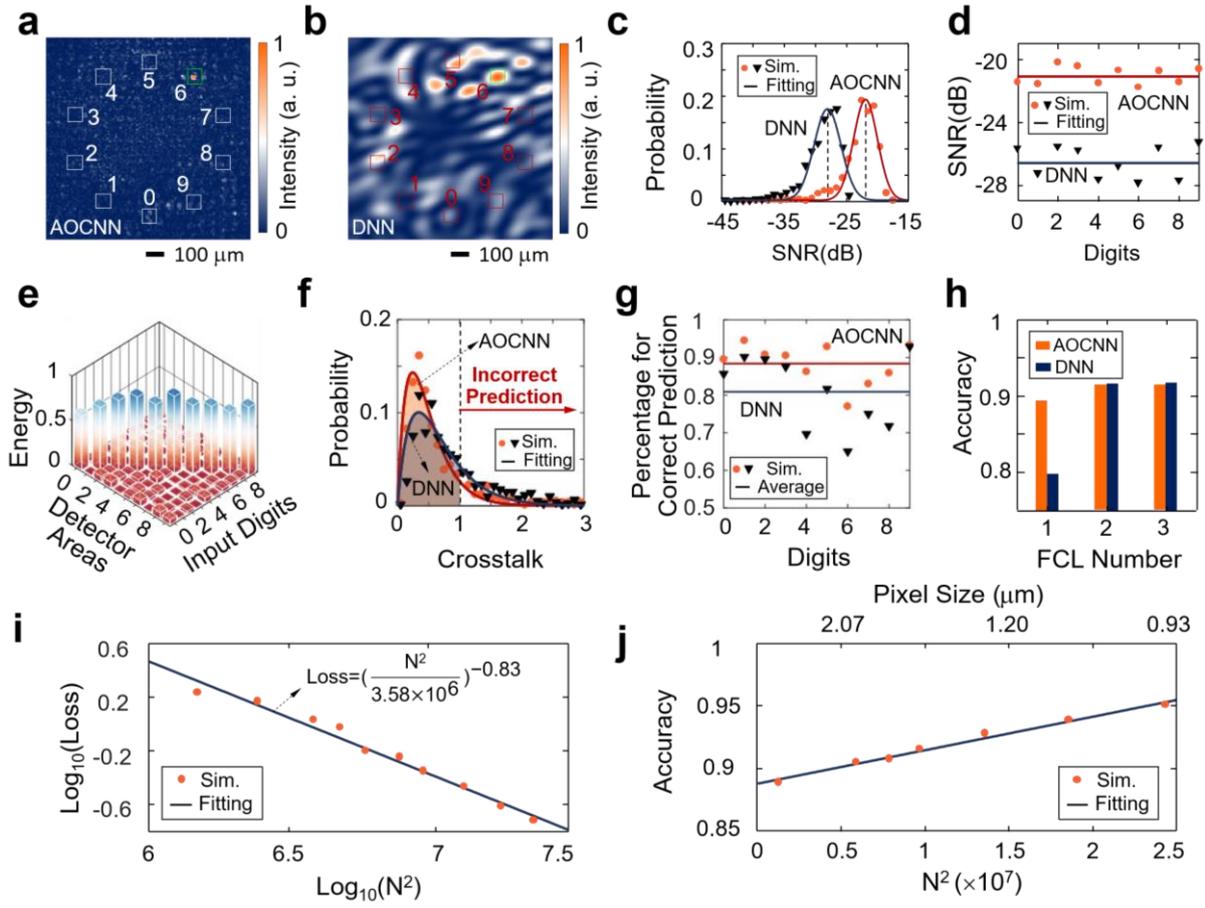

**Figure 2 | Optical performance of AOCNN. a-b,** Simulated images at the detector plane by using AOCNN **(a)** and single-layer DNN **(b)** for the exemplified input digit '6'. **c,** Statistic distributions of calculated signal-to-noise ratios (SNRs) for 800 randomly selected digits "6" from MNIST dataset by using AOCNN (dots) and single-layer DNN (triangles). In this work, $SNR=10\ log_{10}(I_{signal}/I_{noise})$, where $I_{signal}$ and $I_{noise}$ are the energy values at the target region (green square in **(a)**) of the digit '6' and the non-target (containing the entire calculation window expect the green square in **(a)**) region, respectively. **d,** SNRs at peak probability for different digits from '0' to '9' by using AOCNN and single-layer DNN. **e,** Normalized detected energy at 10 detector areas for different input digits for the AOCNN. For each input digit, the normalization is implemented by dividing the energy at each detector region by the total energy from all 10 detector regions. **f,** Simulated (discrete markers) and fitted (solid curves) probability density distribution of the crosstalk when identifying 800 test digits '6' via AOCNN (orange) and single-layer DNN (black). These networks achieve correct recognition only when the crosstalk $C$<1. **g,** Percentage for correct prediction of different digits by using AOCNN (orange dots) and single-layer DNN (dark triangles). **h,** Simulated MNIST classification accuracy with respect to the number of diffractive layers by using AOCNN (orange) and DNN (dark). **i,**

Simulated losses (dots) of training single-FCL AOCNN under different numerical sampling $N^2$. In this simulation, the sampling $N^2$ varies by changing the pixel size of the FCL component and keeping its total spatial area fixed. Fitting result (line) yields a power-law relationship as $Loss=[N^2/(3.58\times10^6)]^{-0.83}$. **j,** Dependence of classification accuracy on the sampling $N^2$ (bottom) or pixel size (top). When the pixel size approaches the wavelength, simulated accuracy can be improved to 95%.

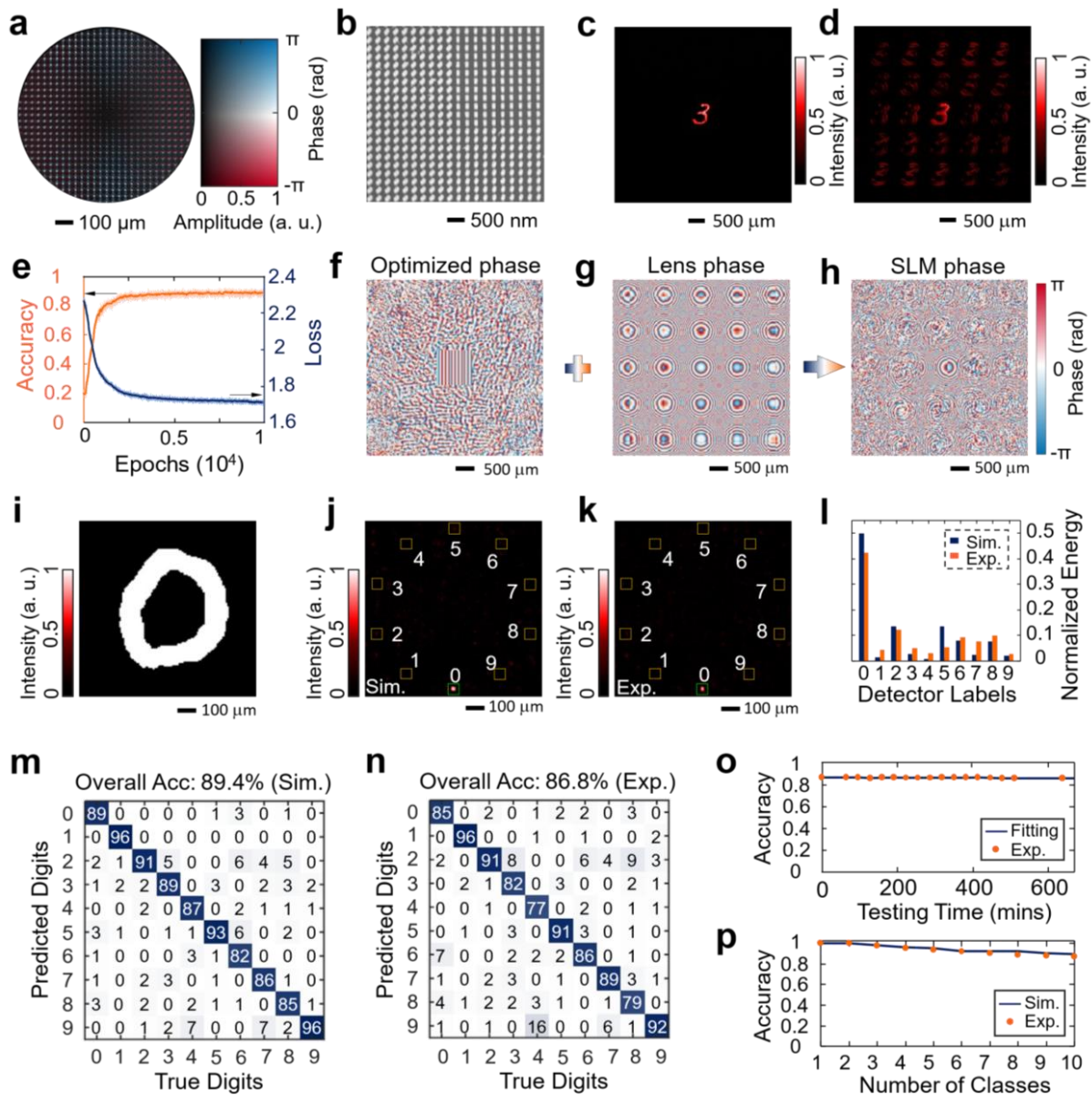

**Figure 3 | AOCNN for digit recognition. a-b,** Complex amplitude profile (**a**) and SEM (**b**) images of our fabricated meta-modulator. **c-d,** Measured images of an input digit '3' for the convolutional layer without (**c**) and with (**d**) the meta-modulator. **e,** Simulated classification accuracy (orange) and loss (dark blue) for the 10-class MNIST test dataset during optimization when the train batch size is 10. **f-h,** Phase profiles for digit recognition. The phase profiles (**h**) loaded on the SLM comprises a trained phase (**f**) combined with a lens

phase (**g**), where the focal length of the lens is 33.33 mm. **i-l,** Recognition result for an exemplified digit '0' (**i**). In the simulated (**j**) and experimental (**k**) intensity profiles on the detector plane, the detector area '0' exhibits significantly higher intensity (**l**) compared to the other nine regions, resulting in the successful recognition of input digit '0'. **m-n,** Simulated (**m**) and experimental (**n**) confusion matrix for evaluating overall accuracies by using our single-layer AOCNN. 100 images for each digit are randomly selected from the test MNIST dataset. **o,** Experimental accuracy (dots) within 10 hours for investigating the stability of our AOCNN setup. **p,** Simulated (solid curve) and experimental accuracy (dots) for MNIST digit classification when the number of classes changes from 1 to 10.

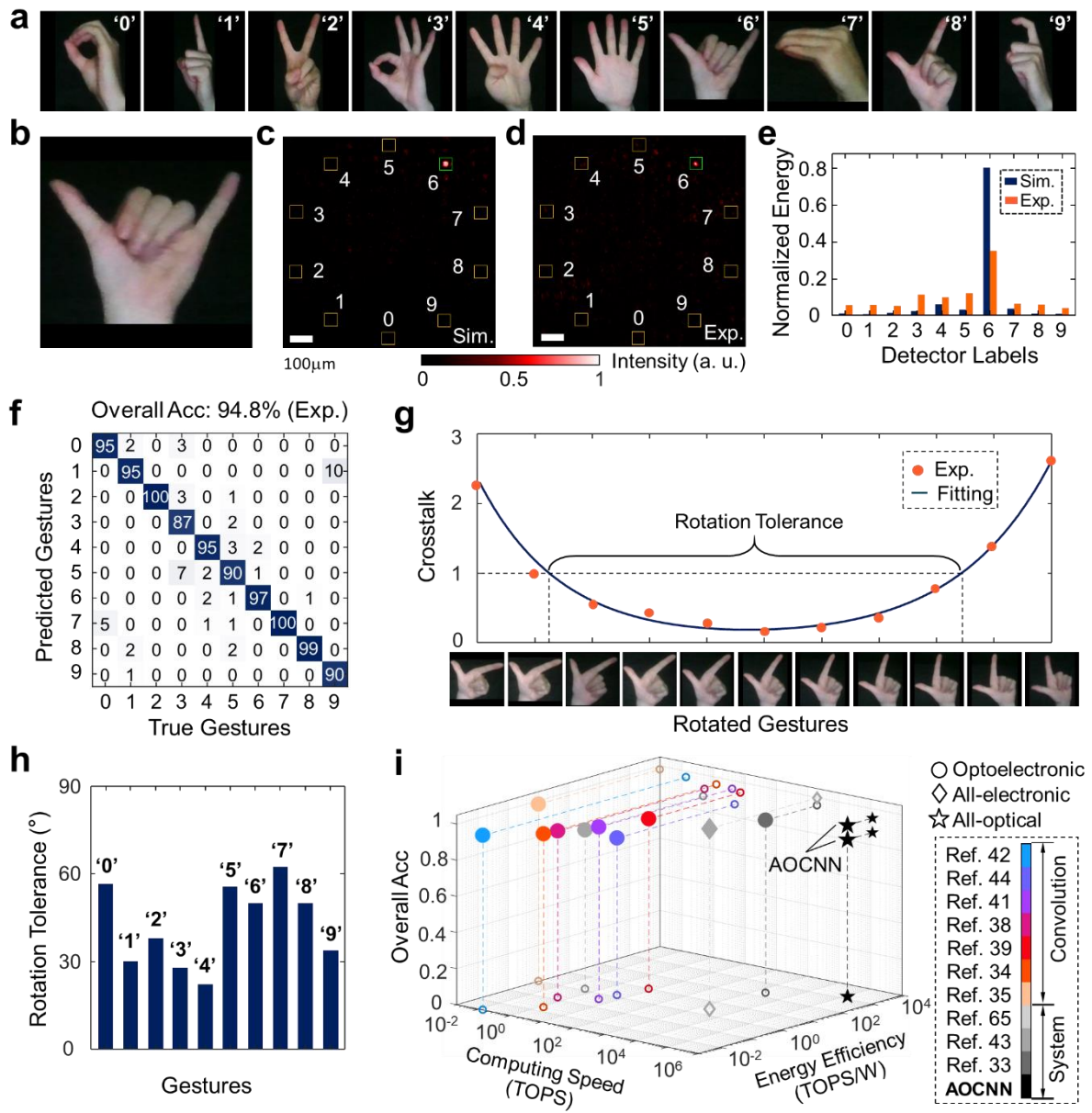

**Figure 4 | Reconfigurable AOCNN for gesture recognition. a,** Gesture images distinguished by their

labels from '0' to '9'. **b-e,** Gesture recognition result for an exemplified gesture '6' **(b)**. In simulated **(c)** and experimental **(d)** intensity profiles, the detector area '6' exhibits much higher energy **(h)** compared to the other nine regions, resulting in successful recognition of gesture '0'. **f,** Experimental confusion matrix. The dataset used in this work consists of 1000 test gesture images (100 per gesture, see Methods). **g,** Simulated and fitted crosstalk for the rotated gesture '8'. The rotation tolerance is specified as the maximum permissible roll angle ensuring that the crosstalk remains below 1. **h,** The experimentally measured rotation tolerance for different gesture categories. The average angle of all ten gesture categories is about 42°. **i,** Computational speed, energy efficiency and accuracy of other previous CNNs trained with MNIST dataset. The works reporting the system efficiency are labelled in gray makers to distinguish from those works (pseudo-color circular dots) with convolution-only efficiency. In addition, different operating modes are also distinguished by using the marker shapes: circle dots for optoelectronic CNNs, diamonds for all-electronic CNNs and stars for all-optical CNNs.

Supplementary Materials for

# An all-optical convolutional neural network for image identification

Wei-Wei Fu[1,2,#], Dong Zhao[1,#], Qing-Hong Rao[1], Heng-Yi Wang[1], Ben-Li Yu[2,3], Zhi-Jia Hu[2,3],

Fang-Wen Sun[1,*], Kun Huang[1,2,*]

[1] Department of Optics and Optical Engineering, University of Science and Technology of China, Hefei, 230026, Anhui, China

[2] State Key Laboratory of Opto-Electronic Information Acquisition and Protection Technology, Anhui University, Hefei, 230601, Anhui, China

[3] School of Optoelectronic Science and Engineering, Anhui University, Hefei, 230601, Anhui, China

[#] *W. F.* and *D. Z.* contributed equally to this work.

[*]Corresponding authors: F. S. (fwsun@ustc.edu.cn), K. H. (huangk17@ustc.edu.cn)

## Table of Contents

**Supplementary Section 1 | Optimization of optical FCLs in AOCNN**

With the fixed convolutional layer architecture, image recognition is accomplished through optimization of the diffraction layer. The optical path structural parameters of the AOCNN comprising $L$ ($L\geq 1$) diffraction layers are illustrated in Supplementary Fig. 1a. As demonstrated in our previous work[1], for input electric field $E_0(x_0, y_0)$, the electric field distribution $E_3$ on the convolutional image array plane can be mathematically represented as:

$$E_3(x_3, y_3) = A\, exp\left[ik\frac{x_3^2+y_3^2}{2(d_2'-f_1)}\right] \cdot E_0(x_0, y_0) \otimes \tilde{P} \otimes \left\{\mathcal{F}\{h_{\text{meta}}(x_2, y_2)\}_{f_x=\frac{x_3+Mx_0}{\lambda(d_2'-f_1)}, f_y=\frac{y_3+My_0}{\lambda(d_2'-f_1)}}\right\}, \quad (S1)$$

where $\tilde{P} = \mathcal{F}\left\{P\left(\frac{x_2+M_1 x_0}{M_2}, \frac{y_2+M_1 y_0}{M_2}\right)\right\} = \tilde{P}(f_x, f_y)|_{f_x=\frac{x_3+Mx_0}{\lambda d_2'}, f_y=\frac{y_3+My_0}{\lambda d_2'}}$ , $P\left(\frac{x_2+M_1 x_0}{M_2}, \frac{y_2+M_1 y_0}{M_2}\right)$ is equivalent aperture of a lens at the $x_2$-$y_2$ plane, $h_{\text{meta}}(x_2, y_2)$ is complex amplitude distribution of the meta-modulator, $\mathcal{F}\{h_{\text{meta}}(x_2, y_2)\}$ is ideal PSF comprising multi-kernel, $M_1 = f_1/d_1'$, $M_2 = (d_2'-f_1)/d_2'$, $M = d_2'/d_1'$ and $A$ is constant.

When the propagation distance $d=0$, the optical field $u_1 = E_3$ maintains its forward propagation, and the corresponding parameters can be mathematically formulated as follows:

$$\begin{cases} n_{i,p}^l = w_{i,p}^l \cdot T_i^l \cdot u_i^l \\ u_i^l = \sum_k n_{k,i}^{l-1} \\ T_i^l = A_i^l \cdot \exp(j\varphi_i^l) \end{cases}, \quad (S2)$$

where $i$ represents the neuron in $l$-th layer, $n_i^l$ represents the output function of the $i$-th neuron in layer $l$-th, $w_i^l$ represents the function of light propagation, $p$ refers to the optical diffraction connection between the $(l+1)$-th layer of neurons and neuron $i$, $u_i^l$ represents the input field of the $i$-th neuron in $l$-th layer, $T_i^l$ represents the transmission coefficient of the $i$-th neuron in $l$-th layer, and $A_i^l$ and $\varphi_i^l$ are amplitude and phase modulations. Since the fully connected layers have only phase modulation, so $A_i^l = 1$. We can obtain loss function:

$$Loss(\varphi_i^l) = -\sum_{n=1}^{N} y_n'^{\mathbf{T}} log\, y_n = -\sum_{n=1}^{N} log\, y_{n,v} = -\sum_{n=1}^{N} log\frac{e^{P_{n,v}}}{\sum_{j=0}^{9} e^{P_{n,j}}}$$

$$= -\sum_{n=1}^{N}(log\, e^{P_{n,v}} - log\sum_{j=0}^{9} e^{P_{n,j}}) = \sum_{n=1}^{N}(log\sum_{j=0}^{9} e^{P_{n,j}} - P_{n,v}), \quad (S3)$$

where $P_k = \sum_{q\in S_k} I_q = \sum_{q\in S_k} |u_q^{L+1}|^2$, $u_q^{L+1} = \sum_k n_{k,q}^L$, $N$ is the number of training samples in a batch, $y_n'^{\mathbf{T}}$ is encoded as a 10×1 One-Hot vector of label, $y_n$ is the normalized matrix of the ten detectors regions energy, $y_{n,v}$ is the normalized energy of the detector '$v$' area energy corresponding to the $n$-th digit '$v$', $P_{n,j}$ is the energy of the detector '$j$' area energy, and $I_q$ is the intensity of point $q$ in the detector '$k$' area. For example, for the $n$-th digit '*v*', only the element at index position *v*+1 in the matrix $\mathbf{y'_n}^\mathbf{T}$ is marked as 1, and the other elements are marked as 0. So, all elements in the matrix $\mathbf{y'_n}^\mathbf{T}$ can be expressed as:

$$y'_{n,v,i} = \begin{cases} 1 & i = v+1 \\ 0 & i \neq v+1 \end{cases}, \tag{S4}$$

where *i* is index position. The FCLs phase parameters are updated using the Adam stochastic gradient descent algorithm[2] to minimize the loss function through error backpropagation (Supplementary Fig. 2b). Consequently, the optimization objective of AOCNN can be formally formulated as the minimization $L(\varphi_i^m)$ ($0 < \varphi_i^m \leq 2\pi$), with the gradient of the loss function derived as follows:

$$\frac{\partial Loss}{\partial \varphi_i^l} = \frac{\partial \sum_{n=1}^{N}(\log \sum_{j=0}^{9} e^{P_{n,j}} - P_{n,v})}{\partial \varphi_i^l} = \sum_{n=1}^{N}\left(\frac{\partial \log \sum_{j=0}^{9} e^{P_{n,j}}}{\partial \varphi_i^l} - \frac{\partial P_{n,v}}{\partial \varphi_i^l}\right)$$

$$= \sum_{n=1}^{N}\left(\frac{\partial \log \sum_{j=0}^{9} e^{P_{n,j}}}{\partial \varphi_i^l} - \frac{\partial P_{n,v}}{\partial \varphi_i^l}\right), \tag{S5}$$

where $\frac{\partial P_k}{\partial \varphi_i^l} = \frac{\partial \sum_{q \in S_k} |u_q^{L+1}|^2}{\partial \varphi_i^l} = 2\sum_{q \in S_k} \text{Real}\left\{(u_q^{L+1})^* \cdot \frac{\partial u_q^{L+1}}{\partial \varphi_i^l}\right\}$. $\frac{\partial u_q^{L+1}}{\partial \varphi_i^l}$ can be written as

$$\frac{\partial u_q^{L+1}}{\partial \varphi_i^{l=L-l'}} = j \cdot T_i^{L-l'} \cdot u_i^{L-l'} \cdot \sum_{k_1} w_{k_1,q}^L \cdot T_{k_1}^L \cdots \sum_{k_{l'}} w_{k_{l'},k_{l'-1}}^{L-l'+1} \cdot T_{k_{l'}}^{L-l'+1} \cdot w_{i,k_{l'}}^{L-l'}, \tag{S6}$$

and $\frac{\partial u_q^{L+1}}{\partial \varphi_i^l}$ can be written as

$$\frac{\partial u_q^{L+1}}{\partial \varphi_i^l} = j \cdot T_i^l \cdot u_i^l \cdot \sum_{k_1} w_{k_1,q}^L \cdot T_{k_1}^L \cdots \sum_{k_{L-l}} w_{k_{L-l},k_{L-l-1}}^{l+1} \cdot T_{k_{L-l}}^{l+1} \cdot w_{i,k_{L-l}}^l, \tag{S7}$$

where *l*<*L*. If *l*=*L*, $\frac{\partial u_q^{L+1}}{\partial \varphi_i^l}$ is updated as:

$$\frac{\partial u_q^{L+1}}{\partial \varphi_i^l} = j \cdot w_{i,q}^l \cdot T_i^l \cdot u_i^l. \tag{S8}$$

The phase $\varphi_{i,t}^l$ at the *t*-th iteration can be updated according to the phase $\varphi_{i,t-1}^l$ at the (*t*-1)-th iteration. Thus, we have phase distribution of *l*-th layer:

$$\varphi_{i,t}^l = \varphi_{i,t-1}^l - \alpha \hat{m}_t / (\sqrt{\hat{v}_t} + \varepsilon), \tag{S9}$$

where $\hat{m}_t = m_t/(1-\beta_1^t)$, $m_t = \beta_1 m_{t-1} + (1-\beta_1)\frac{\partial Loss}{\partial \varphi_{i,t-1}^l}$, $\hat{v}_t = v_t/(1-\beta_2^t)$ and $v_t = \beta_2 v_{t-1} + (1-\beta_2)(\frac{\partial Loss}{\partial \varphi_{i,t-1}^l})^2$. $m_t$ and $v_t$ are the first-order moment (mean) and second-order moment (variance of uncertainty) of the gradient, respectively. The hyperparameter $\beta_1$ denotes the exponential decay rate governing the moving average of gradient moments, where typical implementations employ $\beta_1 = 0.9$ as the default value to balance historical momentum and current gradient contributions. $\beta_2$ is also the exponential decay rate, which controls the influence of the square of the gradient, and the default value is 0.999. In addition, a small constant $\varepsilon=10^{-8}$ is introduced to prevent division by zero, while the learning rate α determines the convergence rate of the optimization procedure.

**Supplementary Section 2 | Introduction to structural similarity index**

The Structural Similarity Index (SSIM) is a perceptual metric that quantifies the similarity between two images[3]. The SSIM index ranges from 0 to 1, where values closer to 1 indicate higher degrees of similarity between the compared images. For given images *x* and *y*, the SSIM can be mathematically expressed as $SSIM(x,y) = [I(x,y)]^\alpha [c(x,y)]^\beta [s(x,y)]^\gamma$, where *I* represents a composite measure incorporating three distinct components: luminance comparison $I(x,y) = \frac{2\mu_x \mu_y + c_1}{\mu_x^2 + \mu_y^2 + c_1}$, contrast comparison $c(x,y) = \frac{2\sigma_x \sigma_y + c_2}{\sigma_x^2 + \sigma_y^2 + c_2}$, and structure comparison $s(x,y) = \frac{\sigma_{xy} + c_3}{\sigma_x \sigma_y + c_3}$, where $\mu_x = \frac{1}{NM}\sum_{i=1}^{N}\sum_{j=1}^{M} x_{ij}$, $\mu_y = \frac{1}{N}\sum_{i=1}^{N}\sum_{j=1}^{N} y_{ij}$, $\sigma_x = \sqrt{\frac{1}{(N-1)(M-1)}\sum_{i=1}^{N}\sum_{j=1}^{M}(x_{ij}-\mu_x)^2}$, $\sigma_y = \sqrt{\frac{1}{(N-1)(M-1)}\sum_{i=1}^{N}\sum_{j=1}^{M}(y_{ij}-\mu_y)^2}$, $\sigma_{xy} = \frac{1}{(N-1)(M-1)}\sum_{i=1}^{N}\sum_{j=1}^{M}(x_{ij}-\mu_x)(y_{ij}-\mu_y)$, and other constants are usually set to : $\alpha = \beta = \gamma = 1$, $c_3 = \frac{c_2}{2}$, $c_1 = (k_1 L)^2$, $c_2 = (k_2 L)^2$, $L=1$, $k_1 = 0.01$, $k_2 = 0.03$.

**Supplementary Section 3 | Optimization of the FCLs in DNN**

The schematic representation of the diffractive neural network (DNN) architecture is presented in Supplementary Fig. 2a. The DNN accepts input images with 228×228 pixels at a pitch of 3.74 μm. The input electric field undergoes *L* diffraction layers with phase modulation, where $d_0$ and $d_l$ (*l*>1) are set to 0 and 85 mm respectively. The electric field energy across ten designated detector regions is computed on the detector plane, with image recognition achieved by identifying the region exhibiting maximum energy. After 10000 training iterations with a batch size of 10, the optimized phase profiles of the DNN with single-layer configuration (*L*=1) are presented in Supplementary Fig. 2b.

**Supplementary Section 4 | Influence of classification accuracy from different factors**

The classification accuracy exhibits a monotonic increase followed by asymptotic plateaus as the number of convolutional kernels grows (see Supplementary Fig. 4a). Consequently, when using the 25 kernels adopted in the main text, further increasing their number yields only marginal gains in accuracy while introducing aberrations due to the expanded field of view in convolutional array imaging plane. Through systematic evaluation of this trade-off, we conclusively employ 25 convolutional kernels as the optimal configuration in our proposed methodology. The classification accuracy also exhibits a progressive improvement as data complexity decreases. Specifically, for tasks where the number of identification categories is no more than 3, the classification accuracy exceeds 97% across the entire range of variations along the *x*-axis. Moreover, we investigate the correlation between detector area dimensions and classification accuracy. Our simulation results (see Supplementary Fig. 4b) demonstrate that reduced detection regions significantly enhance recognition precision. In our proposed configuration, an optimal detector length of *L*=18.7 μm is selected, which achieves superior classification accuracy while simultaneously maintaining tolerance for substantial working distance variations and relaxing the stability requirements for detector alignment.

**Supplementary Section 5 | Dielectric nano-bricks in geometric metasurfaces**

To achieve desired phase and amplitude modulation in our meta-modulator, we employed a dielectric geometric metasurfaces architecture comprising rotationally-tuned silicon nanobricks on a sapphire substrate. The individual nanobrick was oriented with their long axis rotated by an angle $\theta$ relative to the *x*-axis (see the unit cell sketched in the inset of Supplementary Figs. 5a and 5b). The Jones matrix of this structure can be expressed as:

$$\boldsymbol{T}(\theta) = \boldsymbol{R}(-\theta)\boldsymbol{T}_0\boldsymbol{R}(\theta) = \begin{bmatrix} \cos\theta & -\sin\theta \\ \sin\theta & \cos\theta \end{bmatrix} \begin{bmatrix} t_x & 0 \\ 0 & t_y \end{bmatrix} \begin{bmatrix} \cos\theta & \sin\theta \\ -\sin\theta & \cos\theta \end{bmatrix}, \quad \text{S(10)}$$

where $\boldsymbol{R}(\theta)$ is the rotation operator, $\boldsymbol{T}_0$ is the Jones matrix of the unrotated nanobricks, $t_x=|t_x|\exp(i\varphi_x)$ and $t_y=|t_y|\exp(i\varphi_y)$ are the transmission of the $E_x$ and $E_y$ components of the incident light, respectively. For incident circularly polarized light with an electric field $\boldsymbol{E}^\sigma = E \cdot (\boldsymbol{e}_x + \sigma i \boldsymbol{e}_y)$ ($\sigma = \pm 1$), the transmitted field $\boldsymbol{E}_{trans}$ is given by:

$$\boldsymbol{E}_{trans} = \boldsymbol{T}(\theta)\boldsymbol{E}^\sigma = E \begin{bmatrix} \cos\theta & -\sin\theta \\ \sin\theta & \cos\theta \end{bmatrix} \begin{bmatrix} t_x & 0 \\ 0 & t_y \end{bmatrix} \begin{bmatrix} \cos\theta & \sin\theta \\ -\sin\theta & \cos\theta \end{bmatrix} \begin{bmatrix} 1 \\ \sigma i \end{bmatrix}$$

$$= \frac{t_x+t_y}{2}\boldsymbol{E}^\sigma + \frac{t_x-t_y}{2}e^{2i\sigma\theta}\boldsymbol{E}^{-\sigma} = \left|\frac{t_x+t_y}{2}\right| \cdot e^{i\varphi_{prop}^\sigma}\boldsymbol{E}^\sigma + \left|\frac{t_x-t_y}{2}\right| \cdot e^{i\varphi_{prop}^{-\sigma}} \cdot e^{2i\sigma\theta}\boldsymbol{E}^{-\sigma}, \quad \text{(S11)}$$

where $\varphi_{prop}^\sigma$ and $\varphi_{prop}^{-\sigma}$ are the propagation phase of the co- and cross-polarized transmitted light, respectively, and $\tan(\varphi_{prop}^{\pm\sigma}) = (|t_x|\sin\varphi_x \pm |t_y|\sin\varphi_y)/(|t_x|\cos\varphi_x \pm |t_y|\cos\varphi_y)$. The cross-polarized transmitted light $\boldsymbol{E}^{-\sigma}$ has the amplitude modulation of $|\frac{t_x-t_y}{2}|$ and the phase modulation of $\varphi_{prop}^{-\sigma} + 2\sigma\theta$. In our work, we exploit the geometry of the nanobricks to control the amplitude. While the propagation phase $\varphi_{prop}^{-\sigma}$ contributes to additional phase modulation, its variation remains negligible within the designed nanobrick dimensions. So, the geometric phase $2\sigma\theta$ (see Supplementary Fig. 5b) is independent on the geometry of the nanobricks.

To numerically evaluate the conversion efficiency, we used the finite-difference time-domain (FDTD) method for full-wave electromagnetic simulations. The computational domain adopted periodic boundary conditions along the *x*- and *y*-axes, while perfectly matched layers (PMLs) were implemented along the propagation direction (*z*-axis) to minimize spurious reflections. Operating at a design wavelength of 635 nm, we adopted a subwavelength periodicity of $P_1$=250 nm along both *x*- and *y*-directions to achieve optimal conversion efficiency. Through a 15×15 identical unit cell arrangement scheme, this configuration yielded the required operational pixel size of *PS*=3.74 μm. For ease of fabrication, we implemented amplitude modulation by fixing the width *W*=90 nm and varying *L*. The simulation shows a large efficiency in the 0-1 range (see Supplementary Fig. 5a), which means that amplitude modulation is sufficient for our meta-modulator. The conversion efficiency has a peak near *L*=160 nm, so the increasing edge can be used for amplitude modulation. Then we exploited the seven-level amplitude (0, 0.05, 0.1, 0.3, 0.6, 0.85 and 1) by using nanobricks with sizes of (*W*=90 nm, *L*=90 nm), (*W*=90 nm, *L*=100 nm), (*W*=90 nm, *L*=110 nm), (*W*=90 nm, *L*=120 nm), (*W*=90 nm, *L*=130 nm), (*W*=90 nm, *L*=140 nm) and (*W*=90 nm, *L*=160 nm).

## Supplementary Section 6 | Flowchart of acquiring gesture identification video

To achieve real-time gesture image processing, we employed the experimental setup illustrated in Supplementary Fig. 7a. Firstly, we used an external digital camera to acquire within the video frames. Upon detecting a gesture within the video frames, the region containing the gesture was cropped and resized to a resolution of 228 × 228. Subsequently, the image was binarized and loaded onto $SLM_1$. The preprocessed input image was then processed through our optimized AOCNN for gesture recognition. The output image, captured by a CMOS camera, was analyzed by computing the energy distribution across ten predefined detector regions. The final gesture classification was determined by identifying the detector region exhibiting the maximum energy intensity. If the workflow remained incomplete, the system continued to process the subsequent gesture image within the video frames. A detailed schematic of the workflow is provided in Supplementary Fig. 10. The real-time gesture image processing was performed using MATLAB software.

## Supplementary Section 7 | Comparison with other CNNs

In the field of computing hardware performance evaluation, the three key metrics are computational performance, energy efficiency, and inference time. Trillions of operations per second (TOPS) are a simplifying metric for measuring the computing performance of artificial intelligence (AI) hardware. Based on Fourier optics, multiple kernel convolution operations in our convolution layer are equivalent to two fast Fourier transform (FFT) operations and one element-wise complex multiplication operation. Each complex multiplication contains 4 real multiplications and 2 real summations, and each complex summation contains 2 real summations. For an $N \times N$ light field, one FFT requires $5N^2\log_2 N^2$ real-valued operations (OPs) and the complex-valued element-wise multiplication of the meta-modulator modulation requires $6N^2$ real-valued operations. Therefore, the total number of operations in the convolution layer is $R_0=10N^2\log_2 N^2+6N^2$ real-valued operations. When the number of input nodes and the number of output neurons are set to $N^2$, the phase modulation of diffraction layers has $N^2$ complex multiplications, each corresponding to $R_1=6N^2$ real-valued operations. The physical diffraction-weighted connection between the input node and the output neuron has $N^4$ complex multiplications and $N^2(N^2-1)$ complex summations, corresponding to $R_2=6N^4+2(N^2-1)N^2=8N^4-2N^2$ real-valued operations. For our AOCNN with $L$ diffraction layers for $d_0=0$ (Supplementary Fig. 2a), the total number of real-valued operations is $R= R_0+LR_1+LR_2=(10N^2\log_2 N^2+6N^2)+6LN^2+ L(8N^4-2N^2)=10N^2\log_2 N^2+ 8LN^4+4LN^2+6N^2$. In our experiments, the parameters $N$ and $L$ are set to 1240 and 1 respectively, so the total number of operations is R= $1.89\times10^{13}$ OPs.

Since the propagation time of light is negligible, the inference time of our AOCNN is mainly limited by the speed of the sensor. The CMOS camera (SH3-108, Shenzhen Sincevision Technology Co., LTD) captures images at a speed of 6000 FPS, resulting in an inference time of about $1.67\times10^{-4}$ s for one inference. So computational performance of our AOCNN is $1.89\times10^{13}$ Ops$\times$ 6000 FPS =113,400 TOPS. In our prototype AOCNN system, the laser power is about 200 mW, the CMOS camera power is about 30 W, the $SLM_1$ power is about 15 W and the $SLM_2$ power is about 30 W. The energy efficiency is 113,400 TOPS /(30 W+ 200 mW+ 30 W+15 W)= 1507.98 TOPS/W.

Moreover, we compared the reported state-of-the-art convolutional neural network approaches both on-chip and in free space (see Table S1). Comparative analysis demonstrates that our all-optical AOCNN system exhibits superior throughput and energy efficiency compared to state-of-the-art CNN implementations. To better illustrate reconfigurability, we define three distinct levels (see Table S1). Level 1 refers to the weight values and network structure. Once the system is fixed, it is difficult to retrain other tasks. Although the network structure of Level 2 is fixed and the functions of each layer are also fixed, changing the weight values can handle different tasks. The neuron weight values and network structure at level 3 can be reconfigured, including the weights and number of layers at each level, to handle more complex tasks. Thus, the analysis indicates that our AOCNN will have better reconfigurability while maintaining ultra-high-speed task processing in the future.

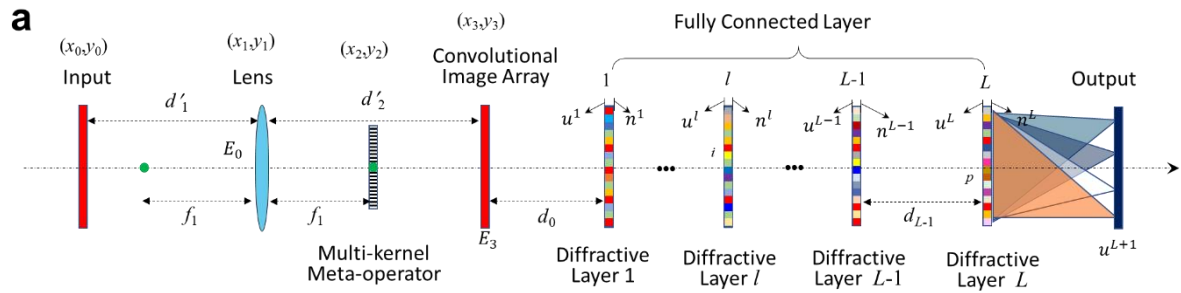

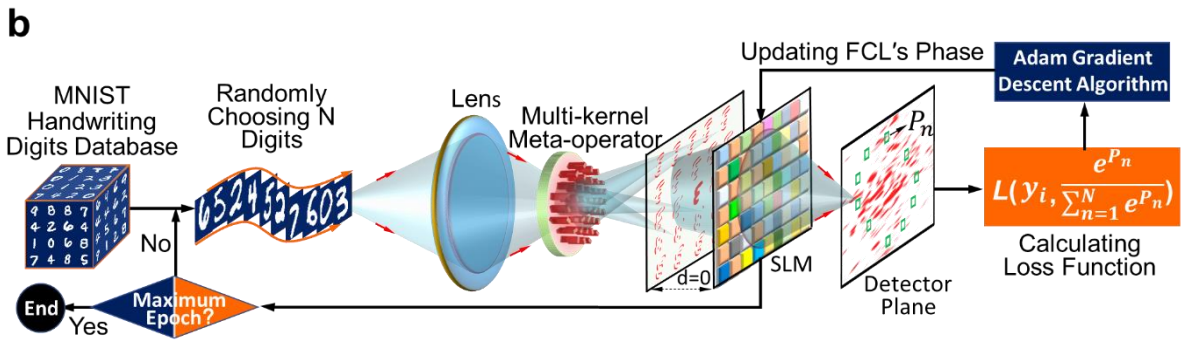

**Supplementary Figure 1 / Optimization of the optical FCLs in AOCNN.**
**a,** Sketch for the AOCNN that contains both optical convolutional layer and optical fully connected layers.
**b,** The process of optimizing AOCNN with a FCL utilizing ADAM gradient descent method.

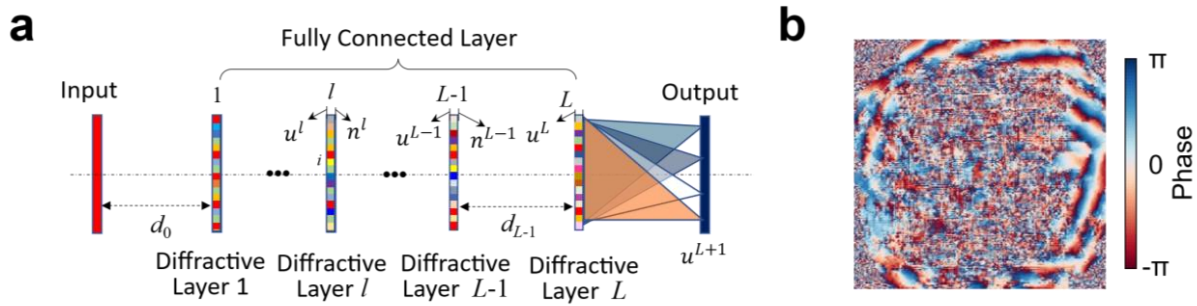

**Supplementary Figure 2 | Design of DNN.**
**a,** Sketch for the DNN that contains L optical fully connected layers. **b,** Optimized phase profiles of DNN with L=1 after 10000 training iterations.

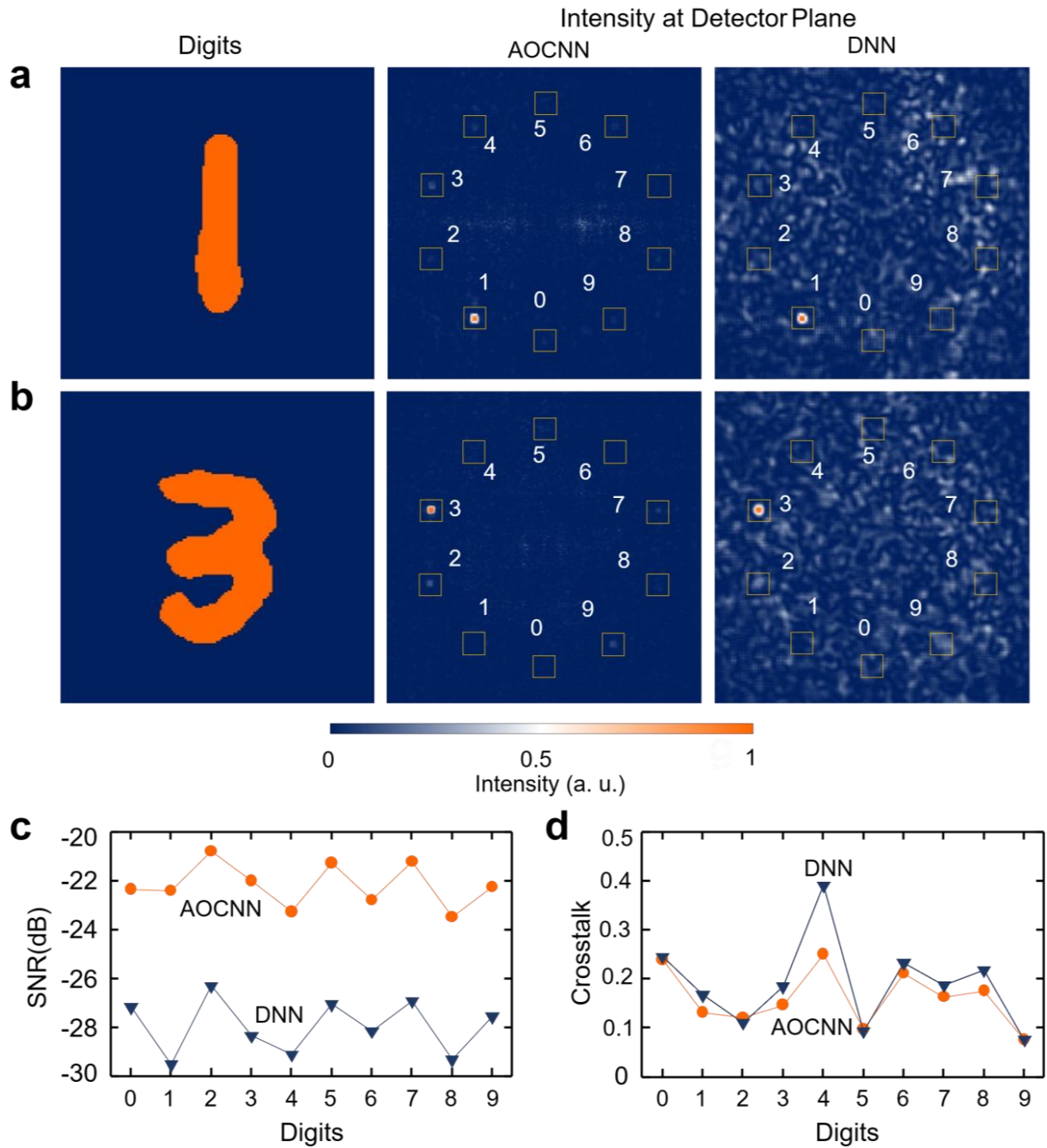

**Supplementary Figure 3 | AOCNN and DNN with two diffractive layers.**
**a,** Simulated intensity distribution at the AOCNN's (middle panel of (**a**)) and DNN's (right panel of (**a**)) detector plane for an input digit '1' (left panel of (**a**)). **b,** Simulated intensity distribution at the AOCNN's (middle panel of (**b**)) and DNN's (right panel of (**b**)) detector plane for an input digit '3' (left panel of (**b**)). **c,** SNR distribution of AOCNN and DNN with different digits. Comparative analysis reveals that the AOCNN achieves a superior SNR, outperforming the DNN by approximately 6 dB. **d,** Crosstalk distribution of AOCNN and DNN with different digits. Both networks exhibit comparable levels.

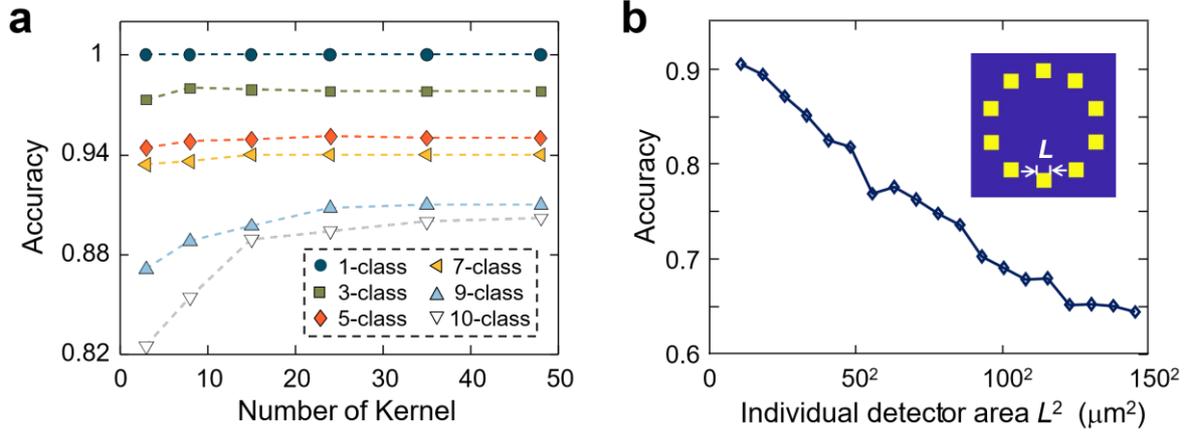

**Supplementary Figure 4 | Simulated MNIST classification accuracy vs the number of convolution kernels, classes and detector regions for a single FCL.**

**a,** Variation of MNIST classification accuracy with respect to the number of convolution kernel and class. To form ideal PSF with different numbers of convolution kernels, the vertical Sobel operator $G_y$ undergoes rotational transformation across a full angular range of 0° to 360° in a manner consistent with Fig. 1b. The total number of rotation angles is the number of convolution kernel minus 1. The center of the ideal PSF is the mean filtering operator, while the rest are Sobel distribution rotated at different angles. **b,** Variation of MNIST classification accuracy with respect to the individual detector area.

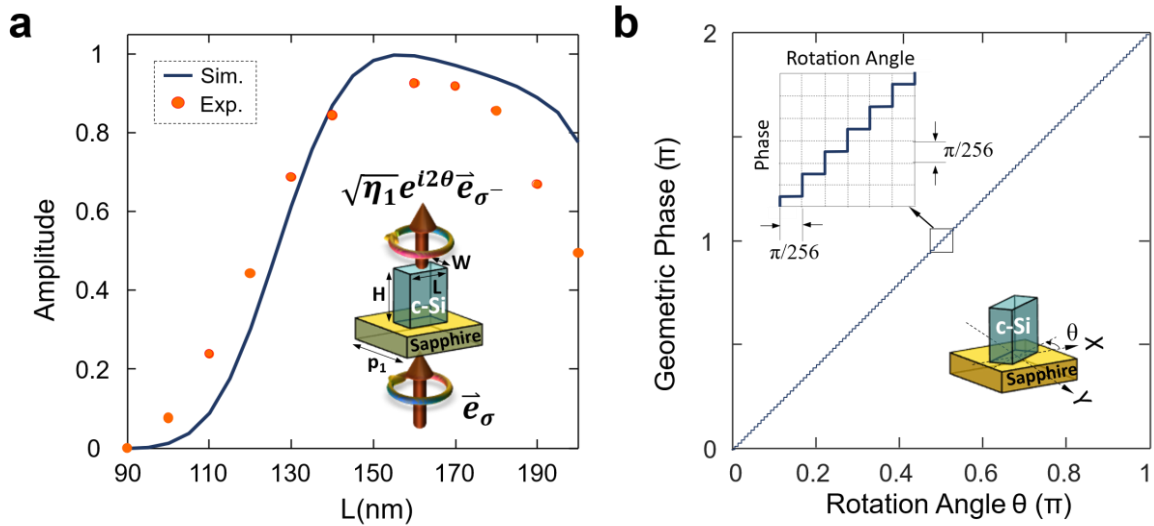

**Supplementary Figure 5 / Design of dielectric geometric metasurfaces.**

**a,** Amplitude modulation when circularly polarized light passes through the nanobricks with different lengths L and fixed widths W. This illustration shows the sketch of unit cell in dieletric geometric metasurfaces. The polarization conversion efficiency is denoted by $\eta_1$. The handedness of circularly polarized light is indicated by σ, whose sign represents the spin state. The spin of circularly polarized light assumes a binary value of ±1, such that two circularly polarized beams with opposite spins are mutually orthogonal. **b,** Phase modulation determined by the rotation angle of nanobricks. To achieve high-precision wavefront control, the meta-modulator incorporates 256 discrete phase modulation levels. The inset schematic depicts the orientation-engineered nanobricks with a defined rotation angle θ.

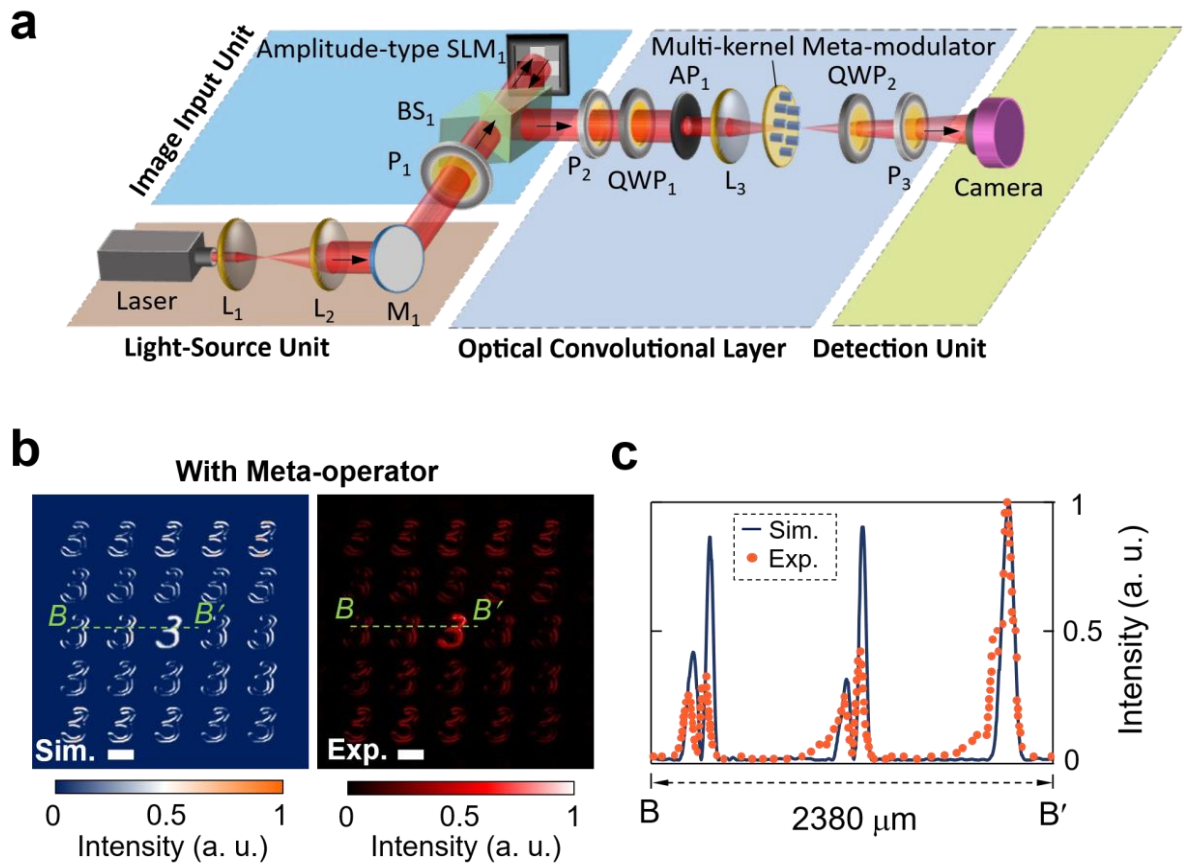

**Supplementary Figure 6 | Experimental characterization of optical convolutional layer.**

**a,** Sketch of the experimental setup for measuring the convolutional image array. Laser: wavelength 635 nm. $L_1$: lens with focal length 100 mm. $L_2$: lens with focal length 25.4 mm. $M_1$: mirror. $BS_1$: beam splitter (50:50). $SLM_1$: spatial light modulator. P: polarizer. QWP: quarter-wave plate. AP: aperture. $L_3$: lens with focal length 50 mm. **b,** Simulated (left panel of (**b**)) and measured (right panel of (**b**)) images by the AOCNN's optical convolutional layer (**a**). Scalebars: 500 μm. **c,** Simulated (curve) and measured (dots) line-scanning intensity profiles along the green line in (**b**).

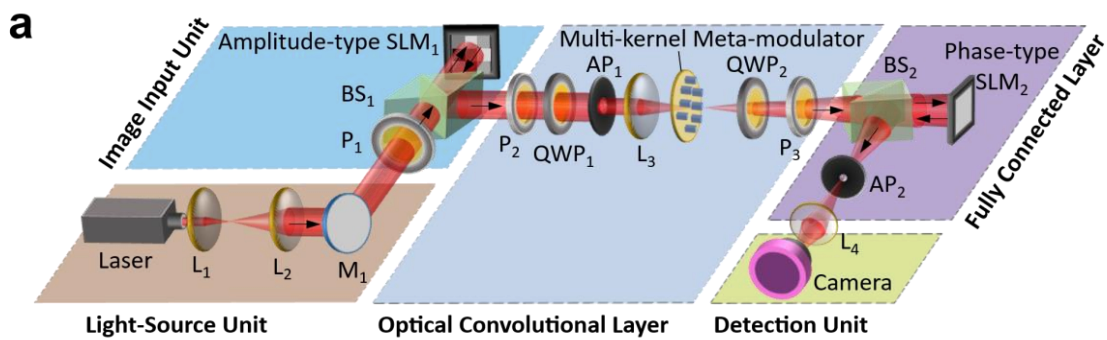

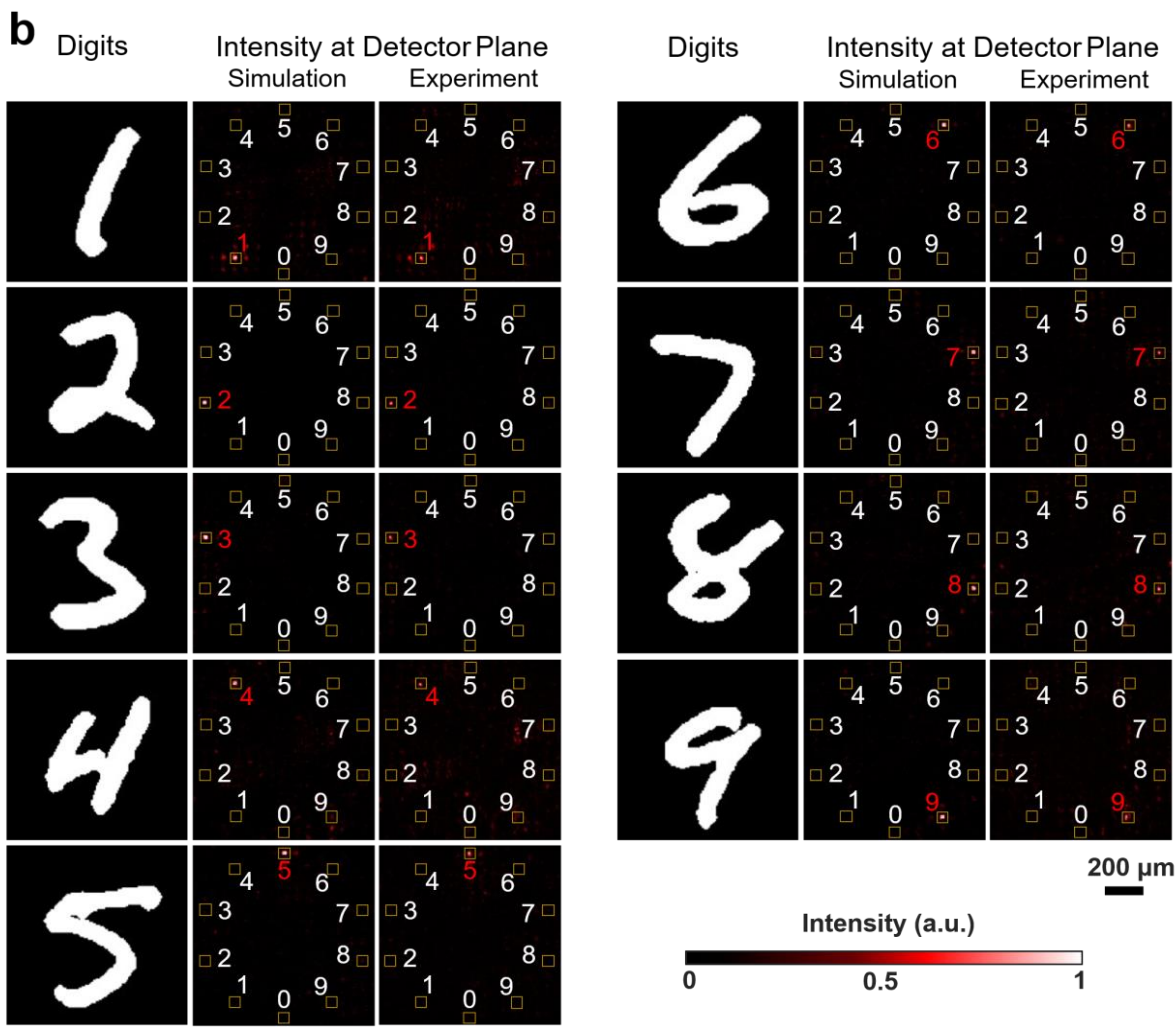

**Supplementary Figure 7 | Experimental characterization of AOCNN.**

**a,** Sketch of the experimental setup for measuring the AOCNN. L$_4$: lens with focal length 100 mm. **b,** Experimental recognition result for inputting numbers '1' to '9' from MNIST dataset. The results demonstrate effective light intensity redistribution to the designated target regions, with experimental measurements showing strong agreement with numerical simulations.

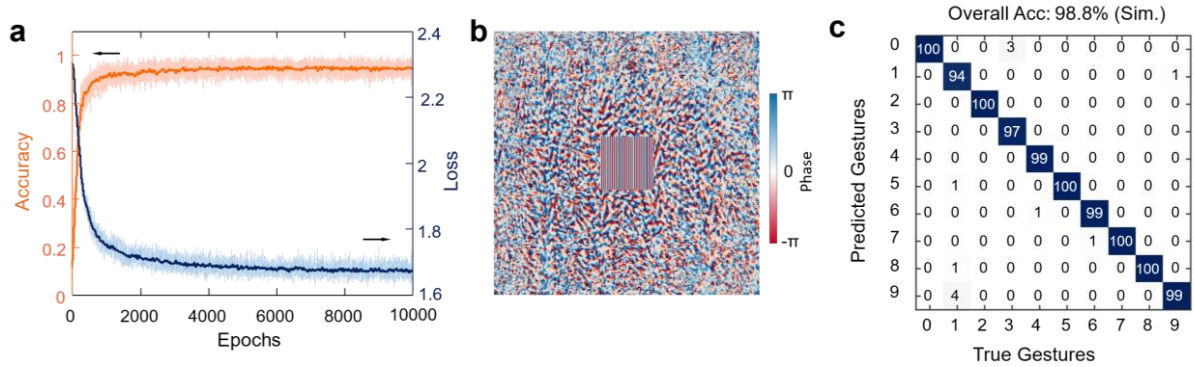

**Supplementary Figure 8 | Training results of the AOCNN with a single FCL for gesture data.**

**a,** Variation of test gesture classification accuracy and loss with respect to the epochs when the train batch size is 10. **b,** Phase profiles of the AOCNN diffraction layer with $L$=1 after 10000 iterations. **c,** Simulated confusion matrix. The dataset used in this work consists of 1000 test gesture images.

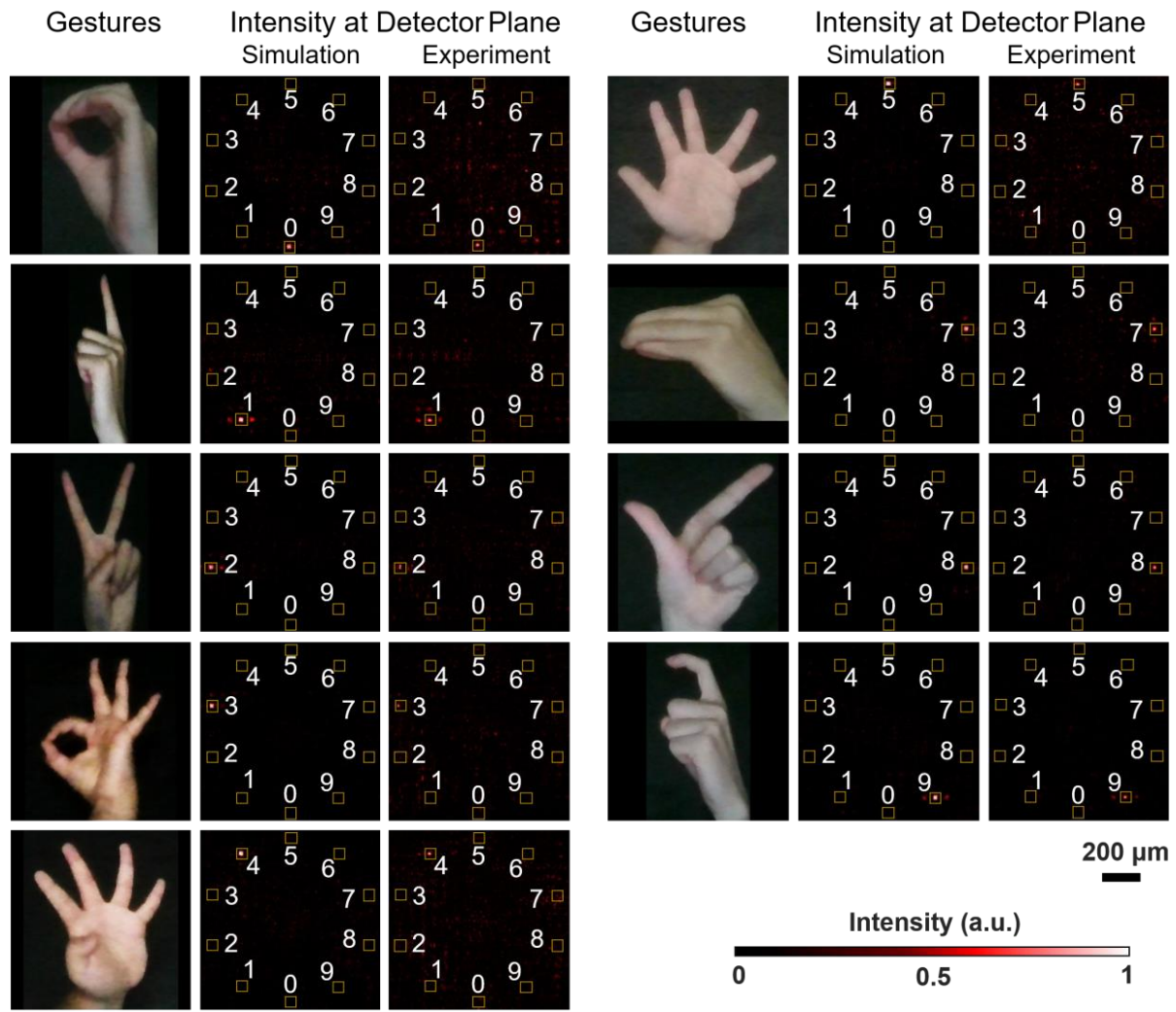

**Supplementary Figure 9 | Experimental gesture recognition using a single-FCL AOCNN.**
The intensity distribution captured at the detector plane after propagation through the trained AOCNN optical path when the input gestures are '0' to '9' except for the gesture '6' from 10-class test dataset (see

Methods). The results demonstrate effective intensity redistribution to the designated target regions, with experimental measurements showing strong agreement with numerical simulations.

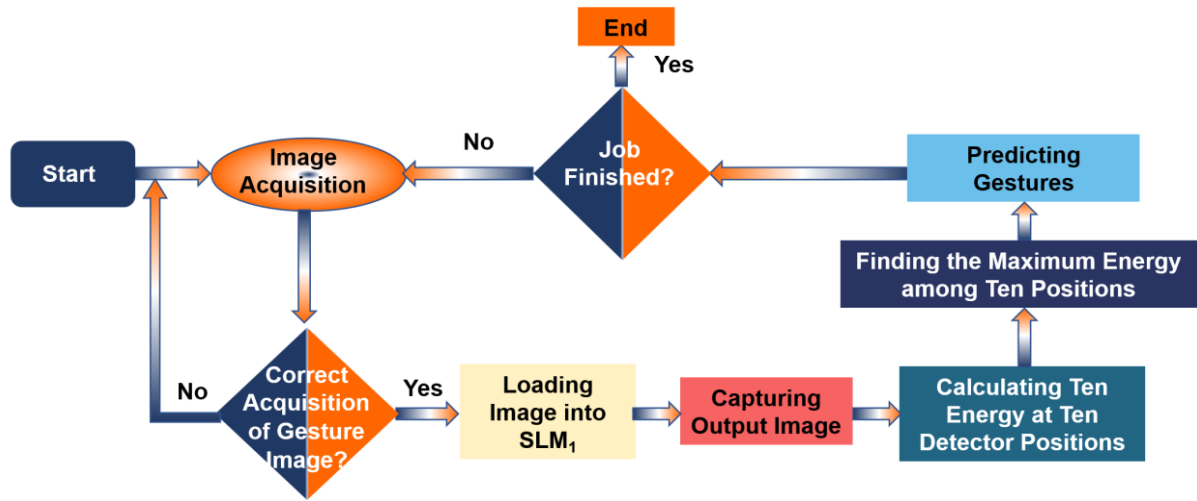

**Supplementary Figure 10 | Optical Flow chart of acquiring gesture identification video**.

**Supplementary Table 11 | A comparison among convolutional neural networks.**

| Working mode | Integration Level | Technology | Reference | Throughout (TOPS) | Energy Efficiency (TOPS/W) | Task modality | Reconfigurability | Accuracy (10-class Handwriting Mnist) |
|---|---|---|---|---|---|---|---|---|
| All-electronic | Yes | 4 nm CMOS chip(GPU H100,NVIDIA) | https://www.nvidia.com/en-us/data-center/h100/ | 1980(System) | 2.83(System) | Multiple | Level 3 | 99.8% |
| Optoelectronic | Yes | MRR array | Nat. Commun. 13, 7970 (2022) | 0.48(Conv.) | 0.06(Conv.) | Multiple | Level 2 | N/A |
| | | MRR array | Nat. Commun. 14, 66 (2023) | 0.136(Conv) | 0.002(Conv.) | Multiple | Level 2 | 96.6% |
| | | Nanobeam array | Optica 11, 190–196 (2024) | 4.8(Conv.) | 1.32(Conv.) | Multiple | Level 2 | 87% |
| | | PCM tensor core | Nature 589, 52–58 (2021) | 4(Conv.) | 0.4(Conv.) | Multiple | Level 2 | 95.3% |
| | | Integrated microcomb | Nature 589,44-51(2021) | 0.48(System) | 1.27(System) | Multiple | Level 3 | 88% |
| | | Programmable waveguide | Nat. Commun. 12, 96 (2021) | 164(Conv.) | N/A | Multiple | Level 2 | N/A |
| | | Spectral filters | Nat. Commun. 16, 81 (2025) | 21(Conv.) | N/A | Multiple | Level 2 | N/A |
| | | WDM and multimode interference | Nat. Commun. 14, 3000 (2023) | 0.53(Conv.) | 0.413(Conv.) | Multiple | Level 2 | 92.17% |
| | | WDM and multimode interference | ACS Photonics. 9, 3906-3916 (2022). | 7.3(Conv.) | 9.5 (Conv.) | Multiple | Level 2 | 94.16% |
| | | MZI mesh and diffraction units | Nat. Commun. 13, 1044 (2022) | 32(Conv.) | N.A.(Conv.) | Multiple | Level 2 | 91.4% |
| | No | 4-f system and optoelectronic weighting elements | Sci. Adv. 9, eadg7904(2023) | 2000(System) | 200(System) | Multiple | Level 2 | 95.6% |
| | | Optical 4-f system and electronic relay layer | Nat. Commun. 14, 7110 (2023) | 1.26(Conv) | 0.02(Conv.) | Multiple | Level 3 | 96% |
| | | 4-f system | Laser Photonics Rev. 16, 2200213 (2022) | 0.02(Conv.) | 1(Conv.) | Multiple | Level 3 | 98% |
| All-optical | No | AOCNN | Our experimental work | $1.13\times10^5$ (System) | 1508 (System) | Multiple | Level 2 | 86.8% |

'*System*' represents end-to-end data, spanning from initial image acquisition to final task execution. '*Conv.*' represents only the convolution layer data during task execution. N/A indicates no available data.